\documentclass[nofootinbib,notitlepage,onecolumn,secnumarabic,amssymb, nobibnotes, aps, prd]{revtex4-1}
\pdfoutput=1 
\usepackage{graphicx,epsfig}
\usepackage{bbold}
\usepackage{amsmath}
\usepackage{ amssymb }
\usepackage{amsfonts}
\usepackage{xcolor}
\usepackage[sort&compress]{natbib}
\usepackage{tikz}
\usetikzlibrary{arrows,positioning,shapes,trees,calc}
\usepackage[export]{adjustbox}
\usepackage{fancyvrb}
\usetikzlibrary{matrix}
\usepackage{array} 
\usepackage{hyperref}

\def\d {\mbox{d}}

\newcommand{\n}{{\bf {n}}}

\newcommand{\obs}{\text{\tiny{obs}}}

\newcommand{\fnl}{{f}_\text{nl}}

\def\k {\bf{k}}
\def\q {\bf{q}}
\def\x {\bf{x}}

\newcommand{\HI}{\text{\tiny{HI}}}
\newcommand{\HH}{\mathcal{H}}

\newcommand{\D} {\partial}

\newcommand{\<}{\langle}
\renewcommand{\>}{\rangle}

\newcommand{\three}{^{\text{\tiny ({{3})}}}}
\newcommand{\two}{^{\text{\tiny ({{2}})}}}
\newcommand{\one}{^{\text{\tiny ({1})}}}

\newcommand{\p}{_{{\text{\tiny$\|$}}}}

\newcommand{\average}[1]{\left\langle #1 \right\rangle_{\tiny{\text{M}}}}

\tikzset{
  treenode/.style = {shape=rectangle, rounded corners,
                     draw, align=center,
                     top color=white, bottom color=blue!20},
  root/.style     = {treenode, font=\Large, bottom color=red!30},
  env/.style      = {treenode, font=\ttfamily\normalsize},
  dummy/.style    = {circle,draw}
}

\begin{document}
\title{Imprint of non-linear effects on HI intensity mapping on large scales}
\author{Obinna Umeh} 
\affiliation{Department of Physics and Astronomy,  University of the Western Cape, Cape Town 7535, South Africa 
}

\date{\today}

\begin{abstract}

Intensity mapping of the HI brightness temperature provides a unique way of tracing large-scale structures of the Universe up to the largest possible scales. This is achieved by using  a low angular resolution radio telescopes to detect emission line from cosmic neutral Hydrogen in the post-reionization Universe.
We use  general relativistic perturbation theory techniques to derive for the first time the full expression for the HI brightness temperature up to third order in perturbation theory without making any plane-parallel approximation. 
We use this result and the renormalization prescription for biased tracers to study the impact of nonlinear effects on the power spectrum of HI brightness temperature both in real and redshift space.
We show how mode coupling at nonlinear order due to nonlinear bias parameters and  redshift space distortion terms modulate the power spectrum on large scales.  The large scale modulation may be understood to be due to the effective bias parameter and effective shot noise. 


\end{abstract}

\maketitle
 \setcounter{footnote}{0}
\DeclareGraphicsRule{.wmf}{bmp}{jpg}{}{}
\maketitle


\section{Introduction}

It  is common in cosmology to assume that only linear perturbation theory is needed for a sufficient description of clustering of Large Scale Structures (LSS) on large scales\cite{Bull:2014rha,Fonseca:2015laa,Alonso:2015uua,Alonso:2015sfa}. 
This is motivated  by the inference drawn from the imprints of LSS on the Cosmic Microwave Background (CMB) radiation\cite{Smoot:1992td}.
There is no evidence yet, to suggest that fluctuations of observed tracers of the underlying  matter density field on large scales is only a linear map  of the  matter density field. 
Rather, evidence from large-scale N-body simulations suggests that simple linear  parametrization of the bias parameter is too simplistic and insufficient to describe clustering of biased tracers\cite{Nishimichi:2011jm}. Semi-analytic treatment of halo clustering  also supports this. It shows that the relationship between haloes and dark matter density field  can be  nonlinear, stochastic and even non-local\cite{Chan:2012jj,Pollack:2013alj}.  
Similarly, it has been shown\cite{Taruya:2010mx} that for an accurate description of damping of Baryon Acoustic Oscillation (BAO)  seen in N-body simulations, nonlinear corrections due to the Redshift Space Distortions, (RSD) must be taken into account.

At the linear order, modes evolve independently, which allows splitting of the RSD effect(especially the part due to gravitationally-induced peculiar velocity) into large-scale Kaiser effect and small scale Finger of God (FoG) effect\cite{Kaiser:1984ApJ}.  Beyond the linear order in cosmological perturbation theory, both long and short wavelength  modes of the peculiar velocity are coupled\cite{Assassi:2014fva}. The mode coupling leads to a modulation of  the amplitude of the long wavelength modes by the short wavelength modes \cite{Assassi:2014fva}. This paper investigates the consequences  of this coupling on the clustering of HI on large scales.

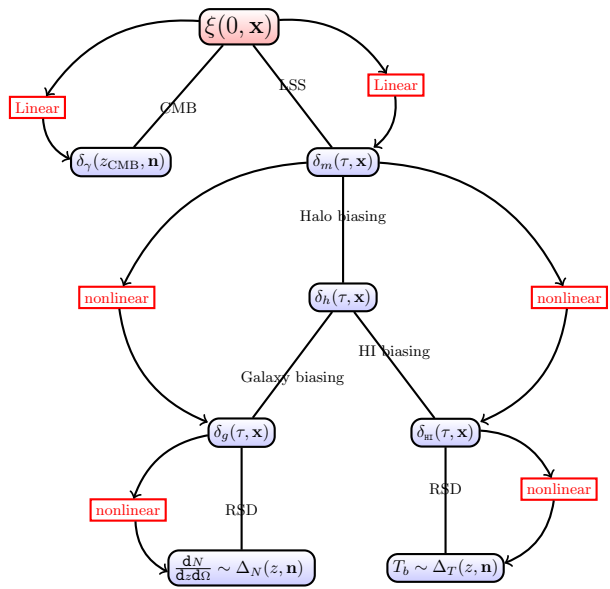
\begin{figure}[htp]
\begin{tikzpicture}[thick,scale=1.2, every node/.style={scale=0.65},pil/.style={
           ->,
           thick,red,
           shorten <=2pt,
           shorten >=2pt,}]
  [
    grow                    = right,
    sibling distance        = 7em,
    level distance          = 12em,
    edge from parent/.style = {draw, -latex},
    every node/.style       = {font=\footnotesize},
    sloped
  ] 
  \node [root](ini) {$\xi(0,{\x})$}
    child { node [left,env](dgamma) {$\delta_{\gamma}(z_{\rm{CMB}},{\n})$}
      edge from parent node [below] (cmb) {CMB} }  
    child { node [right,env](dm) {$\delta_m(\tau, {\x})$}
      child { node [env] (dh) {$\delta_h(\tau, {\x})$}
        child { node [left,env] (dg) {$\delta_g(\tau, {\x})$}
        child { node [env](dN) {${\frac{\d N}{\d z\d \Omega} }\sim \Delta_N(z, {\n}) \,\,$}
        edge from parent node [below] (rsdg) {RSD}}
          edge from parent node [below] {Galaxy biasing} 
           }
        child { node [right,env](dHI) {$\delta_{\HI}(\tau, {\x})$}
        child { node [env] (Tb){$T_b \sim \Delta_T(z, {\n})$}
        edge from parent node [above] (rsdHI) {RSD}}
                edge from parent node [above] {HI biasing} }
        edge from parent node [above] {Halo biasing} }
              edge from parent node [above](lss) {LSS} };

         \node[red,draw, right = 2.5cm of {$(dh)$}] (pertHI) {nonlinear};
         \node[red,draw, left = 2.5cm of {$(dh)$}] (pertg) {nonlinear};
         \node[red,draw, right =  of {$(rsdHI)$}] (pertrsdHI) {nonlinear};
         \node[red,draw, left =  of {$(rsdg)$}] (pertrsdg) {nonlinear};
   
      edge[pil, bend left=45] (dm.east)
       \draw[->] (pertHI) to[bend left] (dHI);
       \draw[<-] (pertHI) to[bend right] (dm);
   
       edge[pil, bend right=45] (dm.west)
        \path[->] (pertg) edge [bend right] (dg);
        \path[<-] (pertg) edge [bend left] (dm);
            
        edge[pil, bend left=45] (dHI.east)
        \path[->] (pertrsdHI) edge [bend left] (Tb);
        \path[<-] (pertrsdHI) edge [bend right] (dHI);
        
        edge[pil, bend right=45] (dg.west)
        \path[->] (pertrsdg) edge [bend right] (dN);
        \path[<-] (pertrsdg) edge [bend left] (dg);
        
        \node[red,draw, right =  of {$(lss)$}] (pertlss) {Linear};
         \node[red,draw, left =  1.5cm of {$(cmb)$}] (pertcmb) {Linear};
         
         edge[pil, bend left=45] (ini.east)
        \path[->](pertlss) edge [bend left] (dm);
        \path[<-] (pertlss) edge [bend right] (ini);
        
         edge[pil, bend right=45] (ini.west)
        \path[->] (pertcmb) edge [bend right] (dgamma);
        \path[<-] (pertcmb) edge [bend left] (ini);
\end{tikzpicture}
\caption{This is a tree diagram showing how observed tracers of the underlying matter density field relate to the initial condition of the universe. A comparison is made with the CMB to illustrate the level of complexity involved.  The red boxes on the sides indicate the level of analysis technique required to sufficiently describe the underlying physical effect. At the top of the tree is the state of the  initial condition that we hope to understand. The CMB and fluctuations in the dark matter density field probe the initial conditions directly and on large scales, the linear theory does capture the underlying physics with sufficient accuracy. The link between the perturbation in dark matter density field and fluctuations of the observed tracers, for example, number count of galaxies or HI brightness temperature, involves two distinct physical processes (i.e halo/galaxy/HI bias and RSD) which cannot be described sufficiently within  the linear theory on any scale. }
\label{fig:one}
\end{figure}

We illustrate in figure \ref{fig:one}, how nonlinearity in the bias and RSD  affect clustering of  observed fluctuations of the HI brightness temperature and galaxy number count. 
Our treatment of these effects is based on  general relativistic perturbation theory, which we use  to derive for the first time the expression for the HI brightness temperature up to third order in perturbation theory. The expression we found includes all important general relativistic corrections at the linear order, RSD and weak gravitational lensing corrections at higher orders.  We then explore in detail how nonlinearity in the bias and RSD contribute to the power spectrum  on large scales.  We show that for the HI power spectrum at one-loop level, nonlinear effects in the local bias model introduce stochasticity in the correlation between HI density fluctuation and the  underlying matter density field. And that it could lead to a spurious scale-dependence in the effective bias parameter on large scales if it isn't properly accounted for. Also, we quantify the error associated with using the linear theory prediction for the HI power spectrum  to analyse observation on large scales

Recent work in this direction includes \cite{Umeh:2015gza}, where it was pointed out that nonlinearity in the bias parameter  modulates the  power spectrum of the HI brightness temperature  on ultra-large scales.  The study was not conclusive because third order perturbation of the HI brightness temperature was not captured in the study, hence not all important one-loop terms were consistently included.  The modification of the galaxy  power spectrum on large scales by nonlinear bias parameters  in real space was studied in \cite{Heavens:1998es} and steps on how the  bias parameters could be re-organized to take into account the effect of mode coupling were discussed in  \cite{McDonald:2006mx,McDonald:2009dh}. This approach was extended in \cite{Assassi:2014fva} to include other two-loop terms. Scoccimarro (2004) \cite{Scoccimarro:2004tg} discussed extensively the validity limits of the standard perturbation theory techniques. He showed that the prediction of the standard perturbation theory in redshift space cannot be trusted below the nonlinear scale. Here, we focus attention  on scales larger than nonlinear scale, where the  suitability of perturbation theory techniques is not under any debate. 

Our fiducial model is determined by the Planck 2015 best-fit values \cite{Ade:2015xua}; in particular,  $h =H_0/100= 0.678$,   $\Omega_{m0} =1-\Omega_{\Lambda 0}= 0.308$. A busy reader can skip section \ref{sec:equation} which covers general relativistic  perturbation theory. Details on how we compute the power spectrum of the HI brightness temperature in redshift space are given section \ref{sec:redshiftspace}. The key results  are presented and discussed in section\ref{sec:resultsanddiscussion}. We conclude in section \ref{sec:conc}. We have three sections in the appendix: In appendix \ref{sec:LOS}, we discuss how to split perturbations into components parallel to the line-of-sight (LOS) and transverse to the LOS. We review details of statistics of dark matter perturbation in Appendix \ref{sec:darkmatter} and explain how we obtain HI bias parameters from halo bias parameters in Appendix \ref{sec:bias}.

\section{Perturbation of HI brightness temperature}\label{sec:equation}

Intensity mapping (IM) of HI brightness temperature is a novel technique capable of mapping the large-scale structures of
the universe in three dimensions. This is achieved by measuring the intensity of
the redshifted 21cm neutral Hydrogen (HI) line over the sky within a determined redshift range without having  to resolve individual galaxies that contain them.
At low redshift, after re-ionization,  most of the HI  are resident in dense gas clouds in galaxies. These clouds of HI emit a unique intensity at a particular frequency determined by the quantum mechanics of electron spin. The brightness temperature ($T^{\rm obs}$) associated with the HI intensity assuming a blackbody radiation  is given by \cite{Hall:2012wd,Alonso:2015uua}
\begin{eqnarray}\label{eq:arbbrightness}
T^{\rm obs}({z},{\n})&=&\frac{3\pi^2}{4} \frac{\hbar^3 A_{10} }{k_{B}E_{21}}\,n_{\HI}({z},{\n})\bigg| \frac{\d{\lambda}}{\d {z}}\bigg|\,,
\end{eqnarray}
where $z $ is the redshift  of the source, $\lambda$ is the affine parameter distance to the source of the HI signal, ${\n}$ is the  direction of the source,    $E_{21}$ is the proper energy of the emitted photons,  $A_{10}= 2.869 \times 10^{-15} {\rm s}^{-1}$ is the emission rate and $n_{\HI}$ is the number density of the HI atoms.
In real space, expanding the $T^{\rm obs}$ in perturbation theory, simply corresponds to expanding only the number density in perturbation theory, since the redshift Jacobian, $J\sim \big| {\d{\lambda}}/{\d {z}}\big|$ is fully determined by the background model:
 \begin{equation}\label{eq:numberdensity}
n_{\HI}({z},{\n})
=\bar{n}_{\HI}({z})\left[1+ \delta_{\HI}\one({z},{\n}) +\frac{1}{2}\delta\two_{\HI}({z},{\n})+\frac{1}{3!}\delta\three_{\HI}({z},{\n}) + \mathcal{O}(\epsilon^4)\right]\,,
 \end{equation}
 where $\delta_{\HI} = (n_{\HI}- \bar{n}_{\HI}) /\bar{n}_{\HI}$ is the density contrast of the HI number density and $\bar{n}_{\HI}$ is mean number density.
Within standard cosmological perturbation theory, we normally assume that the mean number density $\<{n}_{\HI}  \>(z) = \bar{n}_{\HI} (z) = \bar{n}_{\HI} ^{\text{FLRW}}(z)$, i.e we assume that $\bar{n}_{\HI} $ is determined by the background FLRW spacetime.  This assumption will be corrected when we relate $\delta_{\HI}$ to the underlying matter density field at nonlinear order.
We have truncated the perturbation theory expansion at the third order since it is all that we will need to be  able to calculate the power spectrum at the one-loop level consistently. In redshift space, equation \eqref{eq:numberdensity} is not enough, we need to also expand $J$ in perturbation theory as well because the observed redshift of a source cannot be accounted for entirely by the Hubble flow. 
We use $\delta z$ to denote the  perturbation in the redshift of the source. The perturbation  is  mainly due to the  Doppler effect, the relative gravitational potential between the observer and the source, the integrated effect due to varying gravitational potential wells along the line of sight and other special relativistic effects. $\delta z$ may be expanded in perturbation theory:
 \begin{eqnarray}\label{eq:obsz}
(1+ z_{obs})=(1+ z_{\text{true}})\left[1 + \delta z\right] =(1+ z_{\text{true}})\left[1 + \delta\one z + \frac{1}{2} \delta\two  z+\frac{1}{3!} \delta\three z\right]\,.
 \end{eqnarray}
Using equation \eqref{eq:obsz} we could derive an expression describing how  the real space position of the source is distorted due to these gravitationally induced effects
\begin{eqnarray}\label{eq:rsd}
s^i({z},{\n}) &=& x^i({z}) + \frac{n^i}{\HH}\bigg\{ \delta\one z + \frac{1}{2}\left[\delta\two z -(\delta\one z)^2 \left(1+ \frac{\HH'_s}{\HH^2_s}\right)\right]
\\ \nonumber &&
 +\frac{1}{  3!}\left[\delta\three z 
 - 3 {\delta\one z\delta\two z} \left( 1 + \frac{\HH'}{\HH^2}\right)
 + 
 {(\delta\one z)^3}\left( 2 + 3 \frac{\HH'}{\HH^2} \left(1 + \frac{\HH'}{\HH^2}\right) - \frac{\HH''}{\HH^3}\right)\right]\bigg\}\,,
\end{eqnarray}
 where $\HH$ is the Hubble parameter in the conformal time coordinate system and $'$ denotes derivatives wrt to the conformal time. 
Beyond linear order in perturbation theory,  we shall focus only on the contribution due to  Doppler effects(i.e gravitationally induces peculiar velocity $v^i$), since it  dominates  over all other effects in $\delta z$. So for $n \ge 2$, we set $\delta^{(n\ge 2)} z \approx \partial\p v^{(n\ge 2)}$.
From equations \eqref{eq:rsd} and \eqref{eq:obsz}, the Jacobian of the transformation between real and redshift space becomes 
\begin{eqnarray}\label{eq:dhatlamdbyz}
J({z},{\n})=   \bigg| \frac{\d{\lambda}}{\d {z}}\bigg|&=& \frac{a({z})^3}{\HH({z})}\bigg\{1+\left[\frac{1}{\HH_s}
     \frac{\d\delta\one z}{\d\lambda}\right]
       +\frac{1}{2}\left[\frac{1}{\HH_s}\frac{\d\delta\two z}{\d\lambda}              
        + 2\left(\frac{1}{\HH_s}\frac{\d\delta\one z}{\d\lambda}\right)^2
        \right]
        +
        \frac{1}{6}\bigg[\frac{1}{\HH}\frac{\d \delta\three z}{\d\lambda}        
        + 6 \left(\frac{1}{\HH}\frac{\d \delta\one z}{\d\lambda}\right)^3
          \\ \nonumber &&
       + \frac{3}{\HH}\frac{\d \delta\two z}{\d\lambda}\left( \frac{2}{\HH}\frac{\d \delta\one z}{\d\lambda} \right)
        \bigg]
        \bigg\}\,.
        \end{eqnarray}
Putting equations (\ref{eq:dhatlamdbyz}) and  \eqref{eq:numberdensity} in equation \eqref{eq:arbbrightness} leads to 
    \begin{eqnarray}\label{eq:brightHI}
T^{\rm obs}(z,{\n})&=&  \bar{T} (z)\Big[1+\Delta_{T}\one(z,{\n})+\frac{1}{2}\Delta\two_{T}(z,{\n})+\frac{1}{6}\Delta\three_{T}(z,{\n})\Big]\,,
\end{eqnarray}
where $\bar{T}$ is the mean HI brightness temperature. On an FLRW background space-time, it is given by
  \begin{eqnarray}\label{eq:backgdeltaTbin}
  \bar{T} (z) &=&\frac{3\pi^2}{4} \frac{\hbar^3 A_{10}}{k_{B}E_{21}}\,
\frac{ \bar{n}_{\HI} (z)a(z)^3}{\HH(z)}\approx  566 h \frac{\Omega_{\HI}(z)}{0.003}(1+z)^2\frac{H_0}{\HH(z)} \,\,\mu {\rm K} \,.
\end{eqnarray}
Here $\Omega_{\HI}$ is the comoving HI mass density
in units of the current critical density and $H_0$ is the Hubble constant today. 
The perturbation of the HI brightness temperature $\Delta_T$ is given by
\begin{eqnarray}\label{eq:FirstorderT}
     \Delta_T\one(z,{\n})&=&\delta_{\HI}\one 
     + \frac{1}{\HH} \frac{\d\delta\one z}{\d \lambda}  \,,
     \\ \label{eq:SecondorderT}
     \Delta_T\two(z,{\n})&=&\delta_{\HI}\two
     + \frac{1}{\HH} \frac{\d\delta\two z}{\d \lambda}   + 2 \left( \frac{1}{\HH}\frac{\d\delta\one z}{\d\lambda}\right)^2     
     + 2\delta_{\HI}\one \left( \frac{1}{\HH}\frac{\d\delta\one z}{\d\lambda}\right)
     +2 \Delta\one x^a\nabla_{a}\delta T\one\,,    
    \\  \label{eq:ThirdorderT}
\Delta_T\three(z,{\n})&=&\delta_{\HI}\three
     + \frac{1}{\HH} \frac{\d\delta\three z}{\d \lambda}  + 6\left(\frac{1}{\HH}\frac{\d\delta\one z}{\d\lambda}\right)^3
     + 3\delta\two_{\HI} \left( 
     \frac{1}{\HH}\frac{\d\delta\one z}{\d\lambda}\right)
     +\delta\one_{\HI} \left[
      6\left(\frac{1}{\HH}\frac{\d\delta\one z}{\d\lambda}\right)^2          + \frac{3}{\HH}\frac{\d\delta\two z }{\d\lambda} \right]
 \\ \nonumber &&    
     + \frac{6}{\HH}\frac{\d\delta \one z}{\d\lambda} \left[ \frac{1}{\HH}\frac{\d\delta\two z}{\d\lambda}
    \right] 
      +3\Delta\one x^a\nabla_{a}\delta\two T      
      +3\left( \Delta\two x^a\nabla_{a}\delta\one T
  +\Delta\one x^a\Delta\one x^b \nabla_{a}\nabla_{b} \delta\one T \right)\,.
\end{eqnarray}
 The last term in equations \eqref{eq:SecondorderT} and last three terms in \eqref{eq:ThirdorderT} appear because we have corrected for the fact that photons do not propagate on the background space-time but rather on the physical space-time and that the line of sight direction gets modified by the presence of inhomogeneities as well. This is the so-called Born and Post-Born corrections respectively. They were implemented as follows:
%
  \begin{eqnarray}
  \delta\one T(\bar{x}^a) &\rightarrow& \delta\one T({x}^a) 
  + \Delta\one x^a\nabla_{a}\delta\one T  + 
  \frac{1}{2} \left( \Delta\two x^a\nabla_{a}\delta\one T
  +\Delta\one x^a\Delta\one x^b \nabla_{a}\nabla_{b} \delta\one T \right)
  \\ 
   \delta\two T(\bar{x}^a) &\rightarrow& \delta\two T({x}^a) 
  + \Delta\one x^a\nabla_{a}\delta\two T
  \end{eqnarray} 
 In our notation, $\delta  T$ is proportional to $\Delta_T $ without the Born and or  post-Born correction. $\nabla_a$ is the background covariant derivative and $\Delta x^a$ denotes the difference between real and redshift space. We will replace equation \eqref{eq:FirstorderT} with the general relativistic version presented in \cite{Umeh:2015gza}. 
 The contributions of the general relativistic terms at nonlinear order are sub-dominant on all scales at the power spectrum level \cite{Biern:2015tpa} and also for the angular bispectrum \cite{DiDio:2015bua}. They are important for the bispectrum \cite{Umeh:2016nuh} but we do not consider it here. The dominant terms from the decomposition of $\nabla_a$ are 
\begin{eqnarray}
\Delta\one x^a\nabla_{a}\delta \one T &\approx &
 \Delta\one x_{\p} \nabla_{\p} \delta\one{T} + \Delta\one x_{\bot}^i\nabla_{\bot i} \delta\one{T} \,,\\
  \Delta\one x^a\Delta\one x^b \nabla_{a}\nabla_{b} \delta \one T  &\approx & (\Delta\one x\p)^2 \nabla\p^2\delta \one T  +2\Delta\one x\p\Delta\one {x_{\bot}}^i\nabla_{\bot i}\nabla\p \delta \one T  +\Delta\one {x_{\bot}}^i\Delta\one {x_{\bot}}^j\nabla_{\bot i} \nabla_{\bot j}{\delta \one T }\,,
\end{eqnarray}
The full version of these  expressions are given in appendix \ref{sec:LOS}.
The component of $\Delta x^a$ parallel to the LOS is denoted by $\Delta x\p$ and the  component orthogonal to the LOS is denoted by $\Delta x^i_{\bot}$. $\nabla\p$ is the derivative along the LOS direction $\nabla\p = n^i \nabla_i \,$  and $\nabla_{\bot i}$ is the screen-space projected angular derivative $
\nabla_{\bot i} = N_i{}^j \nabla_j  = \nabla_i - n_i \nabla\p$, where $N_{ij}$ is the metric on the screen space. We have retained $\Delta x^i_{\bot}$ for completeness. In the plane-parallel limit, it is set to zero. The contribution of this component  has not been considered in any study up to third order in perturbation theory.  Given a metric in Poisson gauge(see appendix \ref{sec:LOS} for further details), we find
 \begin{eqnarray}
 \Delta\one x^i_{\bot}&=&\int^{\chi_s}_{0}{(\chi-\chi_s)}\nabla_{\bot}^i\Phi_A\one\d\chi\,,
 \label{eq:bendinganglefirst}\\ 
\Delta\two x_{\bot}^i&=& 
\int_{0}^{\chi_s} ({\chi} - \chi_s)\bigg[\nabla_{\bot}^i\Phi\two_A +2 \nabla_{\bot j}\nabla^i_{\bot}\Phi_A\one(\chi)\Delta\one_{\bot}x^i(\chi) \bigg]\d\chi\,,
\label{eq:bendinganglesecond}
 \end{eqnarray}
 where $\Phi_A = (\Phi +\Psi)$, $\Phi$ and $\Psi$ are the gravitational potential and scalar curvature perturbations respectively. $\chi$ is the comoving distance to the source, it is related to the affine parameter(or conformal time) on the conformal background spacetime, $ \chi = \lambda_o -\lambda = \eta_o - \eta$. Equation \eqref{eq:bendinganglesecond} is in agreement with the argument in \cite{Nielsen:2016ldx} on how to correctly implement the post-Born corrections.
Equation \eqref{eq:bendinganglefirst} is related to the  gravitational lensing bending angle at linear order, while  equation \eqref{eq:bendinganglesecond} is the corresponding expression at second order. For the  component of $\Delta x^a$ along $n^a$ we find
\begin{eqnarray}
\Delta\one x\p &\approx&\frac{1}{\HH_s}\delta\one z = 
  -\frac{1}{\HH_s}\nabla_{\p}v_{s}\one \,,\label{eq:LOS1} 
  \\ 
  \Delta\two x\p
  &=&-\frac{1}{\HH_s}\nabla_{\p}v_{s}\two  + 2\left[\frac{1}{\HH_s^2}\nabla_{\p}v_{s}\one\nabla_{\p}^2v_{s}\one - \frac{1}{\HH_s}\nabla_{\bot i}\nabla_{\p}v_{s}\int^{\chi_s}_{0}\frac{(\chi-\chi_s)}{\chi_s}\nabla_{\bot}^i\Phi_A\one\d\chi\right] \,.
  \label{eq:LOS2}
\end{eqnarray}
  In addition to the  approximation we made for the redshift perturbation at higher order, we shall further assume that the contribution from the  LOS derivative of the peculiar velocity term is greater than its conformal time derivative($\D{\p}{v\one}' \ll \D\p^2v\one$),  hence the derivative of the  redshift perturbation with respect to the affine parameter is approximated as follows ($\d/\d\lambda = \nabla_{\eta} - \nabla\p$):
  \begin{eqnarray}
\frac{\d\delta\one z}{\d\lambda}=\D{\p}{v\one}'-\D\p^2v\one\approx -\nabla^2_{\p}v\one\,,\qquad \frac{\d\delta\two z}{\d\lambda}\approx -\nabla^2_{\p}v\two \,,\qquad \frac{\d\delta\three z}{\d\lambda}\approx -\nabla^2_{\p}v\three\,.
\end{eqnarray} 
 Putting equations (\ref{eq:LOS1}), (\ref{eq:LOS2}), (\ref{eq:bendinganglefirst}) and \eqref{eq:bendinganglesecond} in equations \eqref{eq:SecondorderT} and \eqref{eq:ThirdorderT}  and performing some algebraic simplification lead to:
  \begin{eqnarray}
         \Delta_{T}\one(z,{\n})&=&\delta_{\HI}  \one-\frac{1}{\HH}\nabla^2_{\p}v\one+
      \frac{1}{\HH}\left(b_e-2\HH-\frac{\HH'}{\HH}\right)\nabla_{\p}v\one+\frac{1}{\HH}{\Psi\one}'-\frac{1}{\HH}\left(b_e-3\HH-\frac{\HH'}{\HH}\right)
     \Phi\one
     \\ \nonumber &&
     -\frac{1}{\HH}\left(b_e-2\HH-\frac{\HH'}{\HH}\right)
\int_{0}^{\chi}\d \chi {\Phi_A\one}')\,, 
\label{eq:DeltaT1}\\
   \Delta\two_{T}(z,{\n})&= &\delta_{\HI}  \two-\frac{1}{\HH}\nabla^2_{\p}v\two -\frac{2}{\HH}\delta_{\HI}\one  \nabla^2_{\p}v\one+2\left(\frac{1}{\HH}\nabla^2_{\p}v\one\right)^2
           -\frac{2}{\HH}\nabla_{\p}v\one\left[\nabla\p \delta_{\HI}\one  -\frac{1}{\HH}\nabla\p^3 v\one \right]
            \\ \nonumber&&
            +2\left[\nabla_{\bot i} \delta_{\HI}\one  -\frac{1}{\HH}\nabla_{\bot i}\nabla\p^2 v\one \right]\int_{0}^{\chi}\d\tilde \chi(\tilde\chi-\chi) \nabla^i_{\bot}\Phi_A\one
            \,,
            \label{eq:DeltaT2}
 \end{eqnarray} 
 where we have replaced the linear order result with the full general relativistic version given in \cite{Umeh:2015gza}. Equations (\ref{eq:DeltaT1}) and (\ref{eq:DeltaT2}) were presented in \cite{Umeh:2015gza} and they are in agreement with \cite{DiDio:2015bua}. The evolution bias is related to the mean number density: 
 \begin{equation}\label{eq:evolbias}
     b_e(z)= -\HH(z)\frac{\d\ln [\bar{n}_{\HI} (1+z)^{-3}]}{\d\ln(1+ z)}.
     \end{equation}
 The third order contribution is given by
 \begin{eqnarray}\label{eq:DeltaT3}
 \Delta_T\three(z,{\n}) &=&\delta_{\HI}\three
     - \frac{1}{\HH} \nabla\p^2 v\three  - 6\left(\frac{1}{\HH}\nabla\p^2 v\one\right)^3
     - 3\delta\two_{\HI} \left( 
     \frac{1}{\HH}\nabla\p^2 v\one\right)
     +\delta\one_{\HI} \left[
      6\left(\frac{1}{\HH}\nabla\p^2 v\one\right)^2          - \frac{3}{\HH}\nabla\p^2 v\two \right]
 \\ \nonumber &&    
     + \frac{6}{\HH}\nabla\p^2 v\one \left[ \frac{1}{\HH}\nabla\p^2 v\two
    \right] 
    -\frac{3}{\HH}\nabla{\p}v\one\left[\nabla\p \delta\two_{\HI} - \frac{1}{\HH} \nabla^3\p v\two - \frac{2}{\HH} \nabla\p \delta\one_{\HI} \nabla^2\p v\one - \frac{2}{\HH} \delta\one_{\HI} \nabla^3\p v
\one + \frac{4}{\HH^2} \nabla^2\p v\one \nabla^3\p v\one \right]
\\ \nonumber &&
+3\left(\frac{1}{\HH_s}\nabla_{\p}v_{s}\one\right)^2 \nabla\p^2 \left[\delta_{\HI}  \one-\frac{1}{\HH}\nabla^2_{\p}v\one\right]-\frac{3}{\HH_s}\nabla_{\p}v_{s}\two\left[\nabla_{\p}\delta_{\HI}  \one-\frac{1}{\HH}\nabla^3_{\p}v\one\right]  +\frac{6}{\HH_s^2}\nabla_{\p}v_{s}\one\nabla_{\p}^2v_{s}\one\left[\nabla_{\p}\delta_{\HI}  \one-\frac{1}{\HH}\nabla^3_{\p}v\one\right]  
\\ \nonumber &&
+ 3\left[\nabla_{\bot i} \delta\two_{\HI} - \frac{1}{\HH} \nabla_{\bot i}\nabla\p^2 v\two - \frac{2}{\HH} \delta\one_{\HI} \nabla_{\bot i}\nabla\p^2 v\one  - \frac{2}{\HH} \nabla_{\bot i} \delta\one_{\HI} \nabla\p^2 v\one + \frac{4}{\HH^2} \nabla_{\bot i} \nabla\p^2 v\one \nabla\p^2 v\one\right]
\\ \nonumber && \times
\int_{0}^{\chi}{(\tilde\chi-\chi)} \nabla^i_{\bot}\Phi\one_A\d\tilde \chi
  +
  3\left[\nabla_{\bot i}\delta_{\HI}  \one-\frac{1}{\HH}\nabla_{\bot i}\nabla^2_{\p}v\one\right]
  \int_{0}^{\chi_s} ({\chi} - \chi_s)\nabla_{\bot}^i\Phi\two_A\d\chi 
\\ \nonumber &&  
    - 6\left[ \frac{1}{\HH_s}\nabla_{\bot i}\nabla_{\p}v_{s}\one\int^{\chi_s}_{0}{(\chi-\chi_s)}\nabla_{\bot}^i\Phi_A\one\d\chi\right] 
     \left[\nabla_{\p}\delta_{\HI}  \one-\frac{1}{\HH}\nabla^3_{\p}v\one\right]   
  +6 \left[\nabla_{\bot i}\delta_{\HI}  \one-\frac{1}{\HH}\nabla_{\bot i}\nabla^2_{\p}v\one\right]
  \\ \nonumber && \times
\int_{0}^{\chi_s} ({\chi} - \chi_s)\bigg[ \nabla_{\bot j}\nabla^i_{\bot}\Phi_A\one(\chi)\Delta\one_{\bot}x^j(\chi) \bigg]\d\chi
 -6\frac{1}{\HH_s}\nabla_{\p}v_{s}\one\nabla_{\bot i}\nabla\p\left[\delta_{\HI}  \one-\frac{1}{\HH}\nabla^2_{\p}v\one\right] \int^{\chi_s}_{0}{(\chi-\chi_s)}\nabla_{\bot}^i\Phi_A\one\d\chi
\\ \nonumber &&
+3\nabla_{\bot i} \nabla_{\bot j}\left[\delta_{\HI}  \one-\frac{1}{\HH}\nabla^2_{\p}v\one\right]\int^{\chi_s}_{0}{(\chi-\chi_s)}\nabla_{\bot}^i\Phi_A\one\d\chi \int^{\chi_s}_{0}{(\chi-\chi_s)}\nabla_{\bot}^j\Phi_A\one\d\chi\,.
 \end{eqnarray}
 Equation (\ref{eq:DeltaT3}) is one of the new key results of this paper. 
%
In the plane-parallel limit, equation (\ref{eq:DeltaT3}) reduces to 
\begin{eqnarray}\label{eq:DeltaTNewt}
 {\Delta_T\three}(z,{\n}) &=&\delta_{\HI}\three
     - \frac{1}{\HH} \nabla\p^2 v\three  - 6\left(\frac{1}{\HH}\nabla\p^2 v\one\right)^3
     - 3\delta\two_{\HI} \left( 
     \frac{1}{\HH}\nabla\p^2 v\one\right)
     +\delta\one_{\HI} \left[
      6\left(\frac{1}{\HH}\nabla\p^2 v\one\right)^2   - \frac{3}{\HH}\nabla\p^2 v\two \right]
 \\ \nonumber &&    
     + \frac{6}{\HH}\nabla\p^2 v\one \left[ \frac{1}{\HH}\nabla\p^2 v\two
    \right] 
    -\frac{3}{\HH}\nabla{\p}v\one\left[\nabla\p \delta\two_{\HI} - \frac{1}{\HH} \nabla^3\p v\two - \frac{2}{\HH} \nabla\p \delta\one_{\HI}- \nabla^2\p v\one - \frac{2}{\HH} \delta\one_{\HI} \nabla^3\p v
\one 
\right. \\ \nonumber &&  \left.
+ \frac{4}{\HH^2} \nabla^2\p v\one \nabla^3\p v\one \right]
+3\left(\frac{1}{\HH_s}\nabla_{\p}v_{s}\one\right)^2 \nabla\p^2 \left[\delta_{\HI}  \one-\frac{1}{\HH}\nabla^2_{\p}v\one\right]-\frac{3}{\HH_s}\nabla_{\p}v_{s}\two\left[\nabla_{\p}\delta_{\HI}  \one-\frac{1}{\HH}\nabla^3_{\p}v\one\right] 
 \\ \nonumber &&
 +\frac{6}{\HH_s^2}\nabla_{\p}v_{s}\one\nabla_{\p}^2v_{s}\one\left[\nabla_{\p}\delta_{\HI}  \one-\frac{1}{\HH}\nabla^3_{\p}v\one\right]\,.
\end{eqnarray}
Equation \eqref{eq:DeltaTNewt} is in agreement with the results of \cite{Scoccimarro:1997st,Heavens:1998es,Scoccimarro:1999ApJ,Bernardeau:2001qr} for the number count of galaxies in the Newtonian limit.  
Finally, equation (\ref{eq:DeltaT3}) is the state of the art for the perturbation of HI brightness temperature in redshift space.

\section{Power spectrum of the HI brightness temperature in redshift space}\label{sec:redshiftspace}

In this section, we compute the power spectrum of the HI brightness temperature using the results derived in section \ref{sec:equation} in the plane-parallel limit. The HI density contrast that appears in equations \eqref{eq:DeltaT1}, \eqref{eq:DeltaT2} and \eqref{eq:DeltaT3} is given in conformal Newtonian gauge, while the concept of bias only makes sense in the frame where the HI brightness temperature is at rest, i.e comoving synchronous gauge. Thus, we, first of all, transform $\delta_{\HI}$ into comoving synchronous gauge and at linear order, we find
\begin{equation}\label{bias}
\delta\one_{\HI}=\delta^{{\rm cs}(1)}_{\HI}+(3\HH-b_e) v.
\end{equation}
Beyond the linear order, the difference between the comoving synchronous gauge and the conformal Newtonian gauge is of the order of terms we have neglected. 

A consistent bias model is needed  to relate $\delta^{{\rm cs}}_{\HI}$ to the matter over-density $\delta_m$. We assume that the initial perturbations are Gaussian, and use an Eulerian local bias model, applied up to third order:
\begin{eqnarray}\label{eq:biasillust2}
\delta_{\HI}({\x})  
 = b_1\delta_{m} + 
\frac{1}{2}{b_2}\left[ (\delta_m)^2 - \sigma^2_{k_S}\right]+\frac{1}{3!} b_3\big(\delta_m\big)^3\,,
     \end{eqnarray}
     where $b_1$, $b_2$ and $b_3$ are HI bias parameters. They are related to the derivative(s) of $n_{\HI}$ wrt $\delta_{m}$.
     The effect of the tidal bias has been neglected since it is sub-dominant in the case of the power spectrum\cite{Baldauf:2012hs}. We have subtracted off $b_2\sigma^2_{k_S}/2$ to ensure that the average of  (\ref{eq:biasillust2}) vanishes:
\begin{equation}\label{eq:reno}
 \< \delta_{\HI} \>= 0\,,\qquad {\rm{where}}\quad \sigma^2_{k_S} = \int_{k_{\text{min}}}^{k_S} {\d^3 k \over (2\pi)^3} P_{m}(k).
\end{equation}
Here $\sigma^2_{k_S}$  is the variance of $\delta_m$, $k_S$ is the small-scale cut-off and $P_{m}$ is the linear matter power spectrum. 
At every order, we replace $\delta_{\HI}^{(n)}$ in equations \eqref{eq:DeltaT1}-\eqref{eq:DeltaT3} with 
\begin{eqnarray}
\delta_{\HI}\one &=& b_1\delta_m\one \,, \\
\delta_{\HI}\two &=&  b_2\left((\delta_m\one)^2 - \sigma^2_{k_{S}}\right) + b_1\delta_m\two\,,
\\
\delta_{\HI}\three &=& b_3(\delta_m\one)^3 +b_1\delta_m\three 
+ 3b_2 \delta_m\one \delta_m\two \,.
\end{eqnarray}
We expand $\delta_\text{m}^{(n)}$ in Fourier space according to 
\begin{eqnarray}
\delta_\text{m}^{(n)}({\k})=\int \frac{d^3k_1}{(2\pi)^3}\ldots \int
\frac{d^3k_{n}}{(2\pi)^3} \delta_\text{m}({\k}_1)\ldots \delta_\text{m}({\k}_{n})F_{n}({\k}_1,\ldots,{\k}_{n})\delta^\text{(D)}({\k}_1+\ldots+{\k}_{n}-{\k})\,,
\end{eqnarray}
where $F_{n}$ is the kernel for the matter perturbation in Fourier space, further details on it is given in the appendix.
 We  relate the velocity field to the dark matter density field in the Newtonian limit \cite{Bernardeau:2001qr}
\begin{eqnarray}
v\one({\k}) &=& \frac{\HH f}{k^2} \delta_m({\k})\,,
\\
v\two({\k}) &=& \frac{\HH}{k^2}f \int \frac{\d^3 k_1}{(2\pi)^3}\frac{\d^3 k_2}{(2\pi)^3}\delta_m({\k}_1) \delta_m({\k}_2)  G_2({\k}_1,{\k}_2)
(2\pi)^3\delta^D\left({\k}_1 + {\k}_2 - {\k}\right)\,,
\\
v\three({\k}) &=& \frac{\HH}{k^2}f \int \frac{\d^3 k_1}{(2\pi)^3}\frac{\d^3 k_2}{(2\pi)^3}\frac{\d^3 k_3}{(2\pi)^3}\delta_m({\k}_1) \delta_m({\k}_2)\delta_m({\k}_3)  G_3({\k}_1,{\k}_2,{\k}_3)
(2\pi)^3\delta^D\left({\k}_1 + {\k}_2 + {\k}_3- {\k}\right)\,,
\end{eqnarray}
where $f$ is the rate of growth of structures, the full expression for $G_2$ and $G_3$ are given in appendix (\ref{sec:darkmatter}). The general relativistic corrections to these terms are of the order of terms we have neglected.
We relate the gravitational potential to dark matter field using the Poisson equation
\begin{eqnarray}
{\Phi} \one({\k}) &=& -\frac{3\HH^2 \Omega_m}{2 k^2} \delta\one_m({\k})
=-\frac{3}{2}\Omega_{m0}(1+z) \left(\frac{H_0}{k}\right)^2\delta_m({\k})\,.
\end{eqnarray}
  And introduce the primordial non-Guassianity in the  linear bias parameter only at linear order by re-mapping the linear  bias parameter \cite{Dalal:2007cu,Verde:2009hy}
 \begin{equation}
  b_1\mapsto b_{1} +\Delta b({k}) = b_1(k)\,, \quad \text{where}\quad \Delta b({k}) =
\frac{2\fnl
}{\alpha({k})}\delta_\text{c}\left(b_{1}-1\right)\,,\quad \alpha(k)=\frac{2 k^2 c^2 D T(k)}{3H_0^2
\Omega_\text{m}}\frac{g(z=0)}{g(z_\infty)}\,.
 \end{equation}
 Here the factor ${g(z=0)}/{g(z_\infty)}$ ensures that $\fnl$ value is given in the CMB convention\cite{Camera:2014bwa}, $\Omega_m$ is a dimensionless density parameter for the matter field, $D$ is the growth function for linear matter perturbation. $T(k)$ is matter transfer function and $H_0$ is the Hubble rate today.
%
%
The observed fractional HI  brightness temperature is defined as:
\begin{eqnarray}\label{eq:brightHI}
\Delta_{T}(z,\n ) = \frac{T^{\text{obs}}(z,\n ) - \<T^{\text{obs}}\>(z)}{\<T^{\text{obs}}\>(z)}.
\end{eqnarray}
The all sky average of $\<\Delta_{T}\>$ is zero by definition, but if we assume that $\<T^{\text{obs}}\>= \bar{T} ^{\text{FLRW}}$, we get a non-zero value. In order to ensure that $\<\Delta_{T}\> = 0$, we have to renormalize the background temperature.  In the observed redshift space, the average is more complicated but in the Newtonian limit, the velocity terms do not contribute, thus it reduces to an average of the physical number density
\begin{eqnarray}\label{eq:expectionvalue}
\<\Delta_{T}\> =  \frac{1}{2}\<\Delta_{T}\two\>
= \frac{1}{2}  b_2\sigma^2_{k_S}\,,
\end{eqnarray}
In order to obtain $\<\Delta_{T}\> = 0$ or to ensure gauge invariance at second order, the mean temperature must be modified  $\bar{T} =  \bar{T}^{\rm{FLRW}} +\bar{T}^{\rm{FLRW}} \<\Delta_{T}\>$ and perturbations of $\Delta_{T}$  in Fourier space then becomes
\begin{eqnarray}
\Delta_{T}({\k}) &=& {\Delta_{T}\one}({\k})+\frac{1}{2}\left[{\Delta_{T}\two}({\k})-\<\Delta_{T}\two\>\delta^{D}({\k})\right]  + \frac{1}{3!}{\Delta_{T}\three}({\k})\,,
\end{eqnarray}
where $\Delta_{T}({\k})$ is the Fourier transform of the renormalized temperature fluctuations
 \begin{eqnarray}
 {\Delta_{T}\one}({\k}) &=&  \mathcal{T}\int\frac{\d^3 k_1}{(2\pi)^3}\mathcal{Z}\one({\k}_1)\delta_m({\k}_1){(2\pi)^3}\delta^D({\k}-{\k}_1)
  \,,\label{eq:FourierDeltaT1}
  \\ 
{\Delta_{T}\two}({\k}) &=& \mathcal{T}\int\frac{\d^3 k_1}{(2\pi)^3}\int \frac{\d^3 k_2}{(2\pi)^3}\delta_\text{m}({\k}_1)\delta_\text{m}({\k}_2)\mathcal{Z}\two({\k}_1,{\k}_2){(2\pi)^3}\delta^D({\k}-{\k}_1-{\k}_2) - \<\Delta_{T}\two\>\delta^{D}({\k})
\,,
\\
{\Delta_{T}\three}({\k}) &=& \mathcal{T}\int\frac{\d^3 k_1}{(2\pi)^3}\int\frac{\d^3 k_2}{(2\pi)^3}\int\frac{\d^3 k_3}{(2\pi)^3}\delta_\text{m}({\k}_1)\delta_\text{m}({\k}_2)\delta_\text{m}({\k}_3)\mathcal{Z}\three({\k}_1,{\k}_2,{\k}_3) {(2\pi)^3}\delta^D({\k}-{\k}_1-{\k}_2-{\k}_3)
\,.
\label{eq:FourierDeltaT2}
\end{eqnarray}
Here $ \mathcal{T}= 1/(1 + \<\Delta_{T}\>)$ is the renormalization factor and $\mathcal{Z}^{(n)}$ are kernels in Fourier space: 
\begin{itemize}
\item The first order kernel is given by 
\begin{eqnarray}\label{eq:firstorderkernel}
\mathcal{Z}\one(k, \mu) = b _1 +f\mu^2 + i\mu \mathcal{B}\frac{\HH}{k}+\mathcal{A}\,\frac{\HH^2}{k^2} \,, 
\end{eqnarray}
where  
\begin{eqnarray}
     \mathcal{A}& = &{f}\left(3-\frac{b_e^R}{\HH} - \frac{3}{2}\Omega_m\right)
     - \frac{3}{2}\left[2-\frac{b_e^R}{\HH}+\frac{\HH'}{\HH^2}\right]
     {\Omega_m },
  \label{eq:kernelA}     
     \\
     \mathcal{B}& =& -{f}\left(2-\frac{b_e^R}{\HH}+\frac{\HH'}{\HH^2}\right)\,.
     \label{eq:kernelB}
     \end{eqnarray}
    The evolution bias is modified due to the modification to the global HI brightness temperature via the nonlinear bias parameter\cite{Umeh:2015gza}:
\begin{eqnarray}\label{renormevo}
      b_e^R   = b_e - \HH\frac{(1+z)}{1+\<{\Delta_{T}}\>} \frac{\d \<\Delta_{T}\>}{ \d z}\,.
\end{eqnarray}
The first two terms in equation \eqref{eq:firstorderkernel} correspond to the Newtonian limit of the full result, it is so-called Kaiser formula\cite{Kaiser:1984ApJ}. The  third term captures the Doppler effects and the last term describes the local general relativistic projection effects in the plane-parallel limit.       
\item The  second order kernel for $T^\obs$ is given by 
\begin{eqnarray}\label{eq:Z2}
\mathcal{Z}\two({\k}_1,{\k}_2) &=& b_{1}F_2({\k}_1,{\k}_2) + \mu^2  fG_2({\k}_1,{\k}_2)
+ {b_2}
+\mathcal{K}_{\text{R}}\two({\k}_1,{\k}_2)\,,
\end{eqnarray}
 where  the third term is the nonlinear bias parameter.  The last term in equation \eqref{eq:Z2} incorporates all the nonlinear RSD terms
      \begin{eqnarray}\label{eq:KRSD}
\mathcal{K}_{\text{R}}\two({\k}_1,{\k}_2)&=&(f\mu k)\left[\frac{\mu_1}{k_1} \left(b_1 + f\mu^2_2\right) + \frac{\mu_2}{k_2}\left(b_1 + f\mu^2_1\right) \right]\,.
\end{eqnarray}
Equation \eqref{eq:KRSD} is describes the effect of the mode coupling between; (1) linear order RSD term and the linear order HI over-density, (2) two linear order RSD terms, (3) post-Born correction terms from the HI over-density field and linear order RSD term. An important feature to note in equation \eqref{eq:KRSD} is that it is tracer dependent.

 \item At third order, the kernel is given by 
 \begin{eqnarray}\label{eq:Z3}
\mathcal{Z}\three({\k}_1,{\k}_2,{\k}_3) &=&b_1F_3^s({\k}_1,{\k}_2,{\k}_3)
+\mu^2  fG_3^s({\k}_1,{\k}_2,{\k}_3)+b_3 
 + 3 \left[b_2 F_2({\k}_1,{\k}_2)\right]^s
+\left[\mathcal{K}_{\text{R}}\three({\k}_1,{\k}_2,{\k}_3)\right]^s\,,
 \end{eqnarray}
 where the superscript `s' on each of the term indicates symmetrization on all label indices. Similarly, $F_3$ and $G_3$ capture the three-point mode coupling for the density field and peculiar velocity field respectively. They represent the effect of tidal gravitational interaction, their explicit forms are given in appendix \ref{sec:darkmatter}. The third and fourth terms are due to nonlinearity in the bias parameter. The last term in equation (\ref{eq:Z3}) is the most interesting, it takes the form
\begin{eqnarray}\label{eq:KRSDthirdorder}
  \mathcal{K}_{\text{R}}\three({\k}_1,{\k}_2,{\k}_3)&=&3(f\mu k)\bigg[\left[b_2 +b_1 F_2({\k}_1,{\k}_2) + f\mu^2_{12}G_2({\k}_1,{\k}_2)\right]\frac{\mu_3}{k_3} + \left(b_1 + \mu_3^2f\right)\bigg[\frac{\mu_{12}}{k_{12}} G_{2}({\k}_1,{\k}_2)
  \\ \nonumber &&
  +\frac{(f\mu k)}{2}\frac{\mu_1}{k_1}\frac{\mu_2}{k_2}\bigg] \bigg]\,,
  \end{eqnarray}
where $\mu \equiv {\k}\cdot {\n}/k$ with ${\k}= {\k}_1 + \cdots {\k}_n$,  and 
$ \mu_{ij} \equiv ({\k}_i+{\k}_j)\cdot{\n}/{k}_{ij} = ({\mu_ik_i + \mu_j k_j})/{k}_{ij}\,,$ with $k_{ij} =|{\k}_i + {\k}_j|$.
Equation \eqref{eq:KRSDthirdorder} is a combination of all possible nonlinear RSD effects. It includes the mode-coupling RSD effect to the nonlinear bias,  density field and velocity field.  The most important feature to note is that this term is bias dependent as well, hence its effective contribution is dependent on the type of tracer. 
\end{itemize}

We define the full redshift space power spectrum for $T\obs$ as 
 \begin{eqnarray} 
\<  {\Delta_{T}}({\k})  {\Delta_{T}}({\k}')\>
&=& {P}_{T}({k},\mu) (2\pi)^3\delta^D({\k}+{\k}') 
 =\bigg[{P}_{T}^{11}({k},\mu) + {P}_{T}^{22}({k},\mu)+{P}_{T}^{13}({k},\mu)\bigg](2\pi)^3\delta^D({\k}+{\k}')\,,
\label{eq:rsdpower}
\end{eqnarray}
where $P_T$ is the total power spectrum of $T\obs$, it splits into the tree-level ${P}_{T}^{11}$ and one-loop ${P}_{T}^{22}+ {P}_{T}^{13}$ components:
\begin{eqnarray}
{P}_{T}^{11}({k},\mu) &=& \bar{T}^2\mathcal{Z}\one(k, \mu){\mathcal{Z}^{*}}\one(k, \mu)P_{m}(k) \,,\label{eq:powerSpec}\\
{P}_{T}^{22}({k},\mu) &=&\frac{\bar{T}^2}{2}\int \frac{\d^3k_1}{(2\pi)^3}\bigg[ b_{1}F_2({\k}_1,{\k}-{\k}_1)
 +  \mu^2 f G_2 ({\k}_1,{\k}-{\k}_1)
+ {b}_2
+
\mathcal{K}_{\text{R}}({\k}_1,{\k}-{\k}_1)\bigg]^2
P_{m}(|{\k}-{\k}_1|)P_{m}({k}_1)\,,\label{eq:PT22}
\\ 
{P}_{T}^{13}({k},\mu) &=&\bar{T}^2\mathcal{Z}\one(k, \mu)\left[b_1P^{13}_{\delta\delta}(k) + \mu^2  fP_{\theta\theta}^{13}(k) +\left[\left(  b_3 + \frac{68}{21}b_2 \right) \sigma^2_{k_s} +\mathcal{I}_{R}(k,\mu)\right]P_m(k)\right] \,,
\label{eq:PT13}
\end{eqnarray}
where  $P_{m}({k})$ is the linear power spectrum for the  matter density field, $P^{13}_{\delta\delta}$ and $P_{\theta\theta}^{13}$ are parts of the one-loop component of the matter density field and velocity field respectively. They are given in appendix \ref{sec:darkmatter}. 
We defined an angle $\mu_k$ as  $\mu_k = {\k}_1\cdot{\k}/k k_1$, so that $\mu_1$ may be expressed in terms of $\mu_k$ using addition theorem in the limit where $k$ is aligned to ${\n}$; $\mu_1 = \mu_k \mu + \sqrt{(1-\mu_k^2)(1-\mu^2)}$, we have set $\mu_3=\mu$.  The angle $\mu_2$ is fixed using the momentum constraint $\mu_2 = (k_3 \mu - k_1\mu_1)/k_2$.
Similarly, $k_2$ is fixed ${\k}_2= {\k}-{\k}_1$, so that $k_2 = k\sqrt{r^2 - 2 r\mu_k-1} = ky$, where $k_1 = k r$, $y =\sqrt{r^2 - 2 r\mu_k-1}$ and $k_3 =k$.
It was shown in \cite{Matsubara:2007wj,Donghui2010} that in this limit, it is possible to simplify the $\mathcal{I}_{R}$ term in equation \eqref{eq:PT13} further as follows:
\begin{eqnarray}\label{eq:curlyIintegral}
\mathcal{I}_{R}(k,\mu)&=&
\frac{1}{4\pi^2}\int \mathrm{d} k_1\int_{-1}^{1}\d\mu_k  k_1^2P_m(k_1)
\mathcal{K}_{\text{R}}\three( k,{\k}_1,-{\k}_1) = \sum_{m,n,i,j} \mu^{2m} f^n b_1^i b_2^j \frac{k^3}{(2\pi)^2}\int \d r P_{m}(kr) B_{mn ij}(r)\,,
\end{eqnarray}
where the integrands $B_{mn ij}$  are given in appendix \ref{sec:darkmatter}. 
Without loss of generality, we dropped the imaginary part of the linear order kernel in the ${P}_{T}^{13}$ expression. In the limit where $b_1 =1$, $b_2 =0$, $b_3=0$, $\bar{T}=1$ and $b_e^R =0$, equations \eqref{eq:powerSpec}, \eqref{eq:PT22} and \eqref{eq:PT13} reduce to a corresponding set of equations for the matter power spectrum in redshift space.

In the limit where all the light-cone projection effects are set zero, equations \eqref{eq:powerSpec}, \eqref{eq:PT22} and \eqref{eq:PT13} reduce to a corresponding set of equations for the HI power  spectrum in real space:
\begin{eqnarray}
{P}_{T}^{11}({k}) &=&\bar{T}^2 b_1^2P_m(k)\,,\label{eq:HIP11}
\\
{P}_{T}^{22}({k}) &=&\frac{\bar{T}^2}{2}\int \frac{\d^3k_1}{(2\pi)^3}\bigg[b_{1}F_2({\k}_1,|{\k} -{\k}_1|)
+ {b}_2
\bigg]^2
P_m(|{\k}-{\k}_1|) P_m(k_1)\,,\label{eq:HIP22}
\\
{P}_{T}^{13}({k})&=&\bar{T}^2b _1\left[b_1P^{13}_{\delta\delta}(k)  +\left(  b_3 + \frac{68}{21}b_2 \right) \sigma^2_{k_S} P_m(k)\right]\,.
 \label{eq:HIP13}
\end{eqnarray} 
Beyond the linear order, the real space limit may also be obtained from equation \eqref{eq:PT22} and \eqref{eq:PT13} by setting $\mu =0$, i.e considering only the transverse component of the power spectrum. For $b_1 =1$, $b_2 =0$, $b_3=0$ and $\bar{T}=1$, we obtain the matter power spectrum in real space. 

\section{Results and Discussion}\label{sec:resultsanddiscussion}

We now quantify  the contribution of nonlinear effects to $P_T$. For this analysis,
we use the HI bias parameters computed from a simple Sheth-Tormen halo mass function\cite{Sheth:1999mn}. The details on how we obtain HI bias parameters from halo bias parameters is given in Appendix \ref{sec:bias}.  
The convolution integral in equation \eqref{eq:PT13} is not well-behaved in the UV especially through the dark matter variance $\sigma^2_{k_S}$. Thus, we insert a hard-cut off at nonlinear scale to regulate the integral $(k_{1}^{\rm min}, k_{1}^{\rm max})= (10^{-4},k_{\rm{nl}})h^{-1}\,$Mpc (i.e $\sigma^2_{k_S} =\sigma^2_{k_{\rm{nl}}})$, where we have set $ k_{\rm nl}=0.2 h(1+z)^{2/(2+n_s)}\, \text{Mpc}^{-1}$ (here $n_s=0.96$ is the primordial spectral index)\cite{Smith:2002dz} otherwise the integral limit in the ${P}_{T}^{22}$ component is given by $(k_{1}^{\rm min}, k_{1}^{\rm max})= (10^{-4},10^4)h^{-1}\,$Mpc\cite{Carlson:2009it}.  

\subsection{Renormalization of the HI power spectrum in real space }\label{sec:renormalization}

\begin{figure*}[h!]
\begin{center}%
\includegraphics[width=85mm,height=70mm]{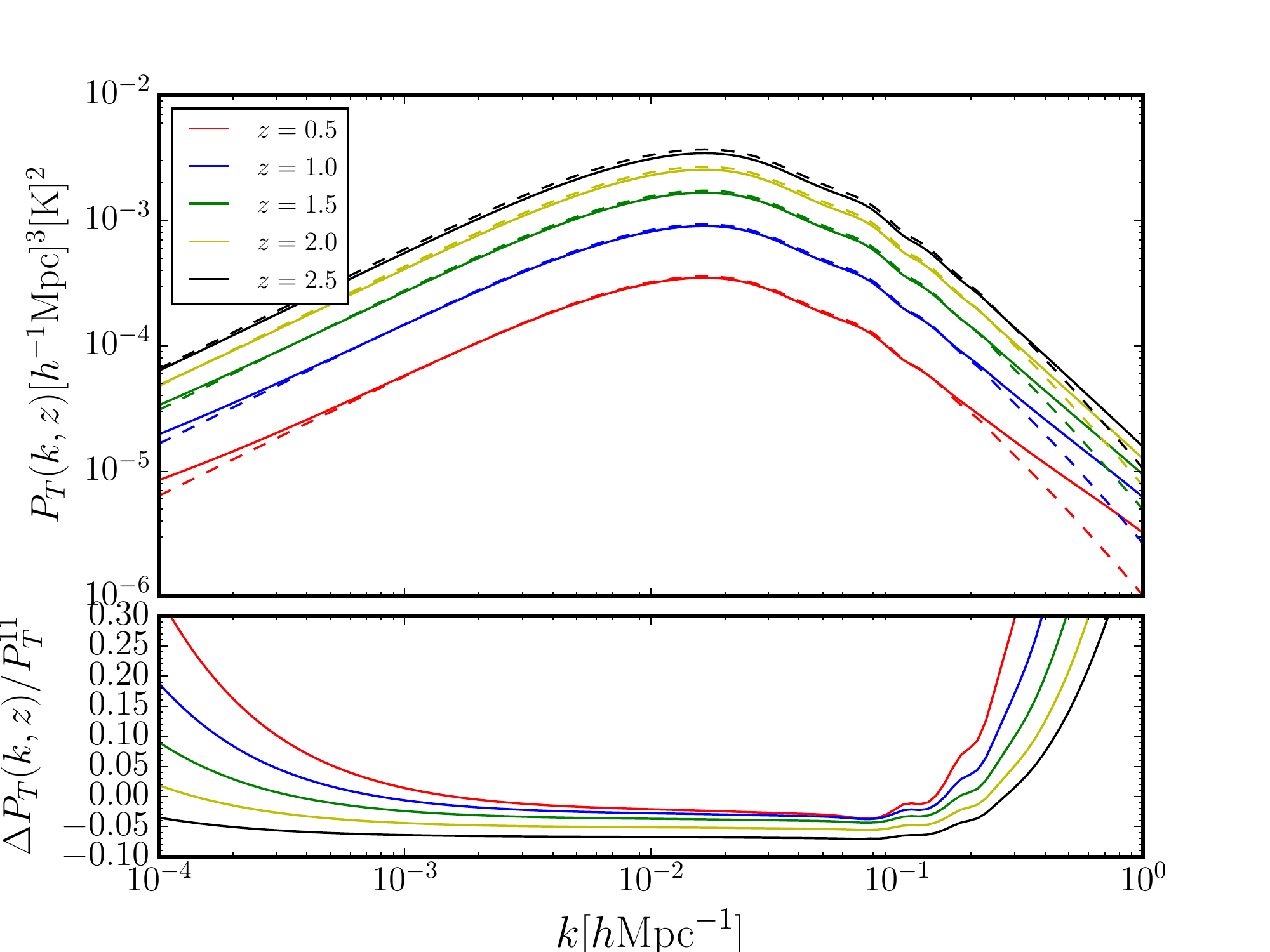}
\includegraphics[width=85mm,height=70mm ]{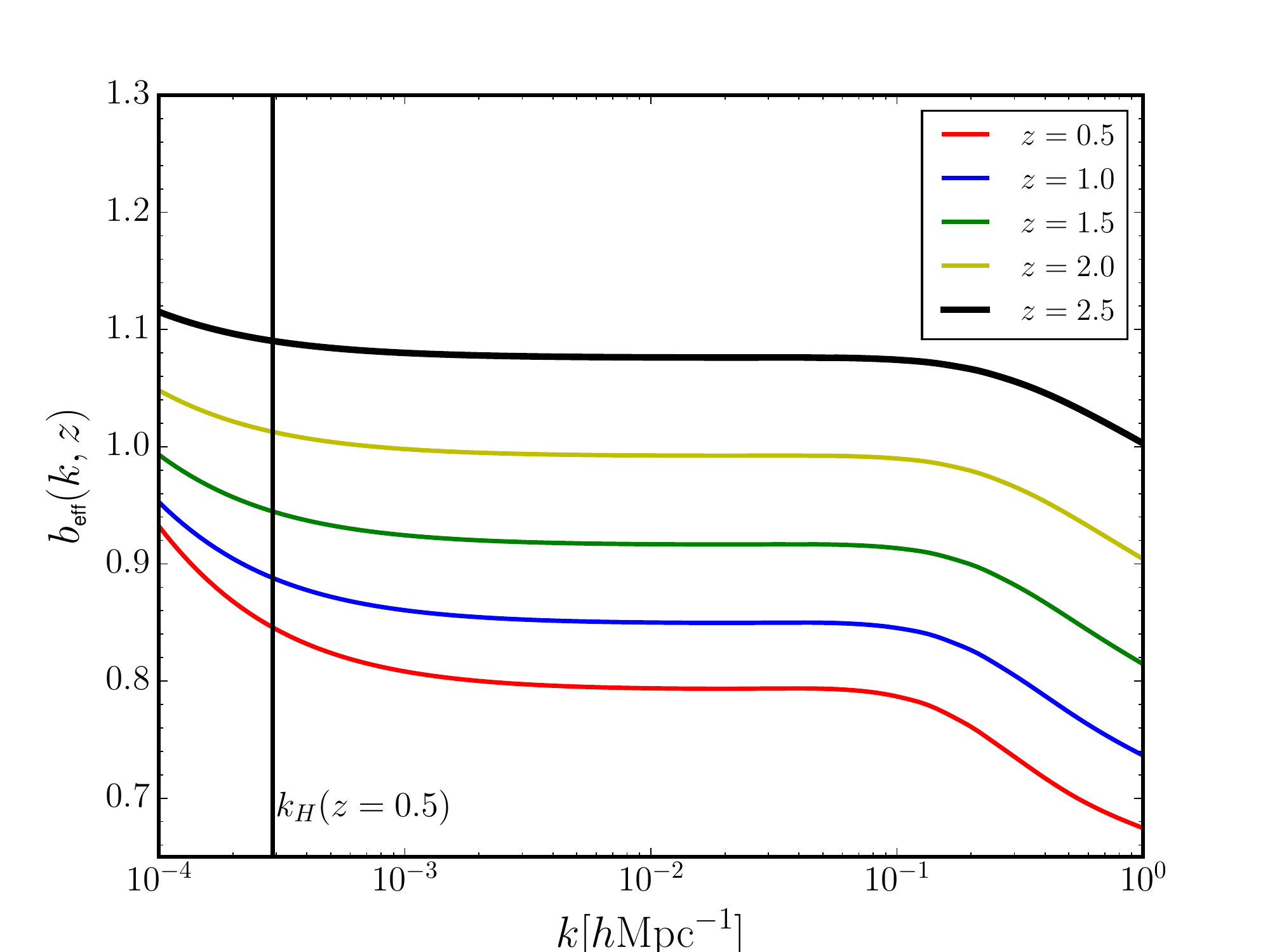}
\caption{ \emph{Left panel}: Real space  $P_T$ at five different redshifts $z = [0.5,1.0,1.5,2.0,2.5]$. Dashed lines are the linear theory result, while the thick lines include all the one-loop terms given in equations \eqref{eq:HIP11}-\eqref{eq:HIP13}. In the lower left panel, we show the fractional difference (${\Delta P_{T}}/{P_T^{11}} = {\big[P_{T} -P_{T}^{11}\big]}/{P_{T}^{11}}$).
\emph{Right panel:} Effective HI bias parameter from the uncorrected  one-loop stochasticity in the auto power spectrum. It is defined in equation \eqref{eq:effectivebias2}. We have normalized $b_{\rm{eff}}$ to remove dependence on $\bar{T}$. The dark vertical line indicates the horizon scale at $z =0.5$.
 }
\label{fig:PkHI1}
\end{center}
\end{figure*}

The HI power spectrum in real space receives corrections from nonlinear bias parameters on all scales.
Most especially, the  $b_2$ term leads to constant power on very large scales and it dominates on super-horizon scales \cite{Heavens:1998es,McDonald:2009dh,BeltranJimenez:2010bb}:
\begin{equation}
k\to 0 ~\Rightarrow~ {P}^{22}_T(k)  \to \frac{1}{2}\left(b_{2}\bar{T}\right)^2  \int \frac{\d^3k_1}{(2\pi)^3}P_{m}^2({k}_1),
\end{equation}
where $F_2({\k},-{\k})=0$. It is possible to renormalize the power spectrum by subtracting this constant power from both sides of  \eqref{eq:HIP22} :
\begin{eqnarray}
{P}_{T}^{22}({k}) &=&\frac{\bar{T}^2}{2}\int \frac{\d^3k_1}{(2\pi)^3}\bigg[\left(b_{1}F_2({\k}_1,|{\k} -{\k}_1|)
+ {b}_2
\right)^2
P_m(|{\k}-{\k}_1|)-\left(b_{2}\right)^2P_m(k_1)\bigg] P_m(k_1) + S_n\,,
\end{eqnarray}
where $S_n$ is constant in $k$ and it behaves 
\begin{eqnarray}\label{eq:defN}
 S_n \equiv \frac{b^2_{2}\bar{T}^2 }{2} \int \frac{\d^3k_1}{(2\pi)^3}{P^2_m}({k}_1)\,.
\end{eqnarray}
 Therefore, the effective $P_{T}$ on large scales is not given by the bare linear theory prediction (i.e ${P}_{T}^{11}(k) = \bar{T}^2b_1^2P_m(k)$) rather it gets a contribution from $P_{T}^{\rm{NL}}$(one-loop correction). The contribution comes from the large scale constant power $S_n$ and the part of ${P}_{T}^{13}$ which do not vanish on large scales $\left(  b_3 + b_2{68}/{21} \right) \sigma^2_{\rm{knl}} P_m(k)$:
\begin{eqnarray}\label{eq:largescalePk}
{P}_{T}^L(k) &\approx &\bar{T}^2\left[b_1 + \frac{1}{2}\left(  b_3 + \frac{68}{21}b_2 \right) \sigma^2_{\rm{knl}}\right]^2P_m(k) +  S_n \,,
\end{eqnarray}
where ${P}_{T}^L(k)$ is the renormalized tree-level power spectrum. 
Note that $S_n$ does not varnish  from the HI power spectrum after renormalization, however, we shall see shortly that renormalization provides an interpretation for the term as stochastic power spectrum. $S_n$ does not contribute to the  cross-power spectrum of the HI over-density and total matter density contrast $\delta $,
\begin{eqnarray}
\<\delta_{\HI}({\k}) \delta({\k}')\> = P_{T\delta}(k)(2\pi)^3 \delta^{D}({\k} + {\k}')\,,
\end{eqnarray}
where $ P_{T\delta}(k)$ is given by 
 \begin{eqnarray}
  \label{eq:HIm}
P_{{T}\delta}(k ) &=&
  \bar{T}\left[b_1 + \frac{1}{2}\left(b_3 + \frac{68}{21}b_2 \right)
   \sigma^2_{k\rm{nl}}\right]P_m(k) 
+\bar{T} b_1\left[P_{\delta\delta}^{13}(k)+P_{\delta\delta}^{22}(k)\right] 
\\ \nonumber && \hspace{1.5in}
+ \bar{T}{b}_2 \int
    \frac{\d^3{k_1}}{(2\pi)^3}\! P_m^{11}(k_1)P_m^{11}(|{\bf{k}-\bf{k}_1}|)
    F_{2}({\bf{k}_1,\bf{k}-\bf{k}_1}).
\end{eqnarray}
On large scales, the effective or renormalized cross-power spectrum  is given by 
\begin{eqnarray}\label{eq:effectivetdelta}
P_{{T}\delta}^{L}(k ) &\approx&
  \bar{T}\left[b_1 + \frac{1}{2}\left(b_3 + \frac{68}{21}b_2 \right)
   \sigma^2_{k\rm{nl}}\right]P_m(k)\,. 
\end{eqnarray}
From equation \eqref{eq:effectivetdelta}  the  effective  bias parameter on large scales is simply a re-parametrization of the linear bias parameter
 \begin{eqnarray}\label{eq:effectivebias}
b_{\text{eff}}^{T \delta}\equiv \frac{P_{{T}\delta}^{L}(k )}{P_{\delta \delta}(k)} \approx b_1 + \frac{1}{2}\left(  b_3 + \frac{68}{21}b_2 \right) \sigma^2_{k\rm{nl}}\,.
\end{eqnarray} 
 However,
if one naively defines the effective HI bias parameter from the auto power spectrum of the HI brightness temperature, it will lead to 
\begin{eqnarray}\label{eq:effectivebias2}
 b_{\rm{eff}}(k) \equiv \sqrt{\frac{P_{T}(k)}{P_{\delta\delta}(k)}} &\approx & \sqrt{\frac{P_{T}^L(k)}{P_{m}(k)}} \sim \bigg\{\left[b_1 + \frac{1}{2}\left(  b_3 + \frac{68}{21}b_2 \right) \sigma^2_{k_{\rm{nl}}}\right]^2 +\frac{S_n}{P_m(k)}\bigg\}^{\frac{1}{2}}\,,
 \end{eqnarray}
 on large scales.
Notice that the effective bias parameter inferred from $P_{T}(k)$ and the effective bias parameter inferred from $P_{{T}\delta}(k)$ differ through $S_n$. The difference denotes the stochasticity in the HI intensity map due to small scale fluctuations\cite{Heavens:1998es}, it is usually described by the cross-correlation coefficient\cite{Hamaus:2010im} and for the local bias model it may be written in this form\cite{Baumann:2012bc}
\begin{eqnarray}
r \equiv \frac{P_{T}(k)}{P_{\delta\delta}(k)} - \left(\frac{P_{{T}\delta}(k)}{P_{\delta\delta}(k)}\right)^2 \approx \frac{S_n}{P_m(k)}\,,
\end{eqnarray}
where we have taken the large scale limit.
This is the randomness that arises whenever a tracer is not 100\% correlated with the underlying matter density field. This indicates that HI can not remain a good tracer of the dark matter density field in the regime where baryon physics plays important role in clustering. Using equation \eqref{eq:effectivebias2} as the effective bias parameter,  this term introduces a spurious scale dependence in the bias parameter on large scales. This is shown in figure\ref{fig:PkHI1}. 

To obtain an effective HI bias parameter from $P_T$, one must quantify this term exactly and then subtract it off from the auto-power spectrum:
$ 
 \sqrt{{(P_{T}(k) -S_n)}/{P_{\delta\delta}(k)}} = b_{\text{eff}}^{T \delta}
$ \cite{McDonald:2006mx,McDonald:2009dh}. It is only after this is done that a correspondence with the effective bias parameter from the cross-power spectrum between HI and matter over-densities(see equation \eqref{eq:effectivebias}) agree.  
Figure\ref{fig:PkHI1} shows that this spurious effect can induce a scale-dependence on the effective bias parameter that could mimic the signature of the primordial non-Gaussinity at low redshift\cite{Camera:2013kpa}.

\subsubsection{Stochastic power spectrum, shot noise and nonlinear effects}

We shall now attempt to link the induced stochasticity due to nonlinear effects on small scales discussed in section \ref{sec:renormalization} to the total noise budget for discrete tracers. 
We follow closely the definition of the stochastic power spectrum for any tracer given in \cite{Baldauf:2016sjb,Baldauf:2013hka,Hamaus:2010im}. This definition assumes that on large scales, over-density is small hence linear local bias model is enough to relate a tracer to the underlying matter density field.
 The stochasticity in the tracer induces a power spectrum which may be defined as 
\begin{eqnarray}\label{eq:stochasticpower}
(2\pi)^3 \delta^{(\text{D})}\left({\k}+ {\k}'\right) C(k) &=& \<\left(\delta_{\HI}({\k}) - b_1\delta_{m} ({\k})\right)
\left(\delta_{\HI}({\k}') - b_1\delta_{m} ({\k}')\right)\>
\\
&=&(2\pi)^3 \delta^{(\text{D})}\left({\k}+ {\k}'\right)\left[P_{T}(k) - 2b_1 P_{_{T\delta}}(k) + b_1^2P_{m}(k)\right]\,,
\end{eqnarray}
where $C(k)$ is the stochastic power spectrum at a given scale $k$, we have set $\bar{T} =1$ in this section in order to reduce clutter. For the local bias model, the linear bias parameter may be defined as $b_1 =P_{T \delta}(k)/P_{m}(k)$ leading to the definition of the \emph{continuous} HI power spectrum  on large scales as
\begin{eqnarray}\label{eq:contpower}
P_{T}^{\text{c}}(k)&=&C(k) +b_1^2P_{m}(k)\,.
\end{eqnarray}
 What is 
 observed/measured is not the \emph{continuous} power spectrum given in equation \eqref{eq:contpower} but the \emph{discrete} tracer power spectrum(HI are resident in galaxies at low z and galaxies are discrete objects). For a finite number of \emph{discrete} sources $N$, at position ${\x}$ in a finite volume $V$, the HI over-density in Fourier space is given by 
 \begin{eqnarray}
 \delta_{\HI}({\k}) = \frac{N}{V} \sum_{i} \exp\left[ i k x_i \right]\,.
 \end{eqnarray}
 The power spectrum associated with the finite number of  discrete sources may be defined as
 \begin{eqnarray}
 P_{T}^{d}(k) &=& \frac{1}{V} \<  \delta_{\HI}({\k} ) \delta_{\HI}({\k}')\> = \frac{V}{N^2}\left[\sum_{i=j}\<\exp\left[ i k (x_i - x_j)\right]\> + \sum_{i\neq j}\<\exp\left[ i k (x_i - x_j)\right]\>\right]
 \\
 &=&P_{T}^{\text{shot}} +   P_{{T}}^{\text{c}}(k) = P_{T}^{\text{shot}}+C(k) + b_1^2P_{m}(k)  = S^{\rm eff}_n  + b_1^2P_{m}(k) \,,
 \label{eq:discretepk}
 \end{eqnarray}
where $P_{T}^{\text{shot}}$ denotes the power spectrum associated with the shot noise and the non-zero separation part is identified with the continuous part of the discrete tracer power spectrum. In the second equality, we have replaced the continuous power spectrum with its definition given in equation\eqref{eq:contpower} and in the third equality, we defined an effective noise parameter as a sum of the stochastic power spectrum and the intrinsic shot noise: $S^{\rm eff}_n = P_{T}^{\text{shot}} +S_n$.  Here, $S_n$ is interpreted as an additional correction to the noise budget due to nonlinear mode coupling\cite{Baldauf:2016sjb}.  This is the rationale behind the identification of $S_n$ as shot noise in\cite{McDonald:2006mx,McDonald:2009dh}.

\begin{figure*}[h!]
\begin{center}%
\includegraphics[width=85mm,height=70mm]{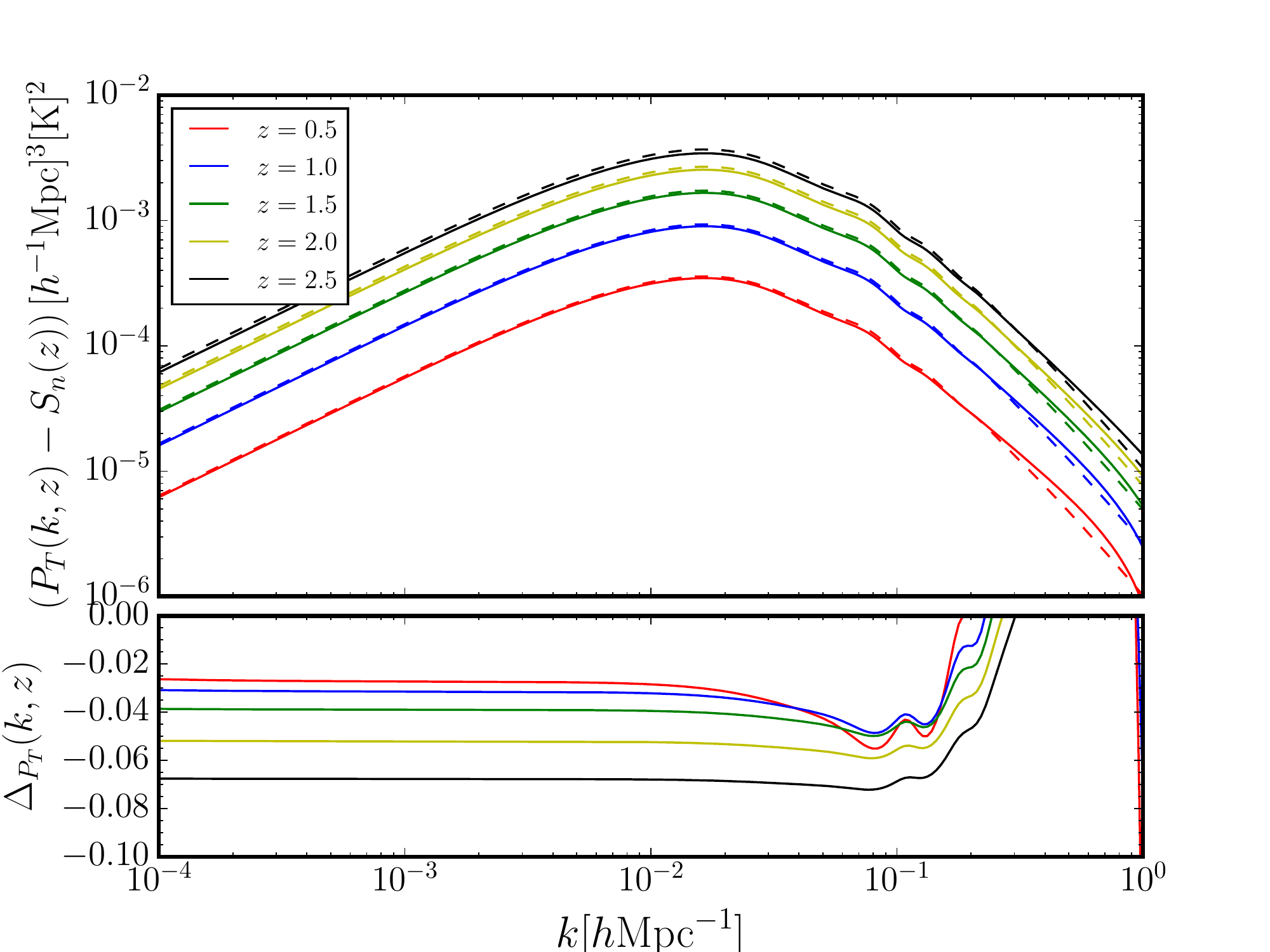}
\includegraphics[width=85mm,height=70mm ]{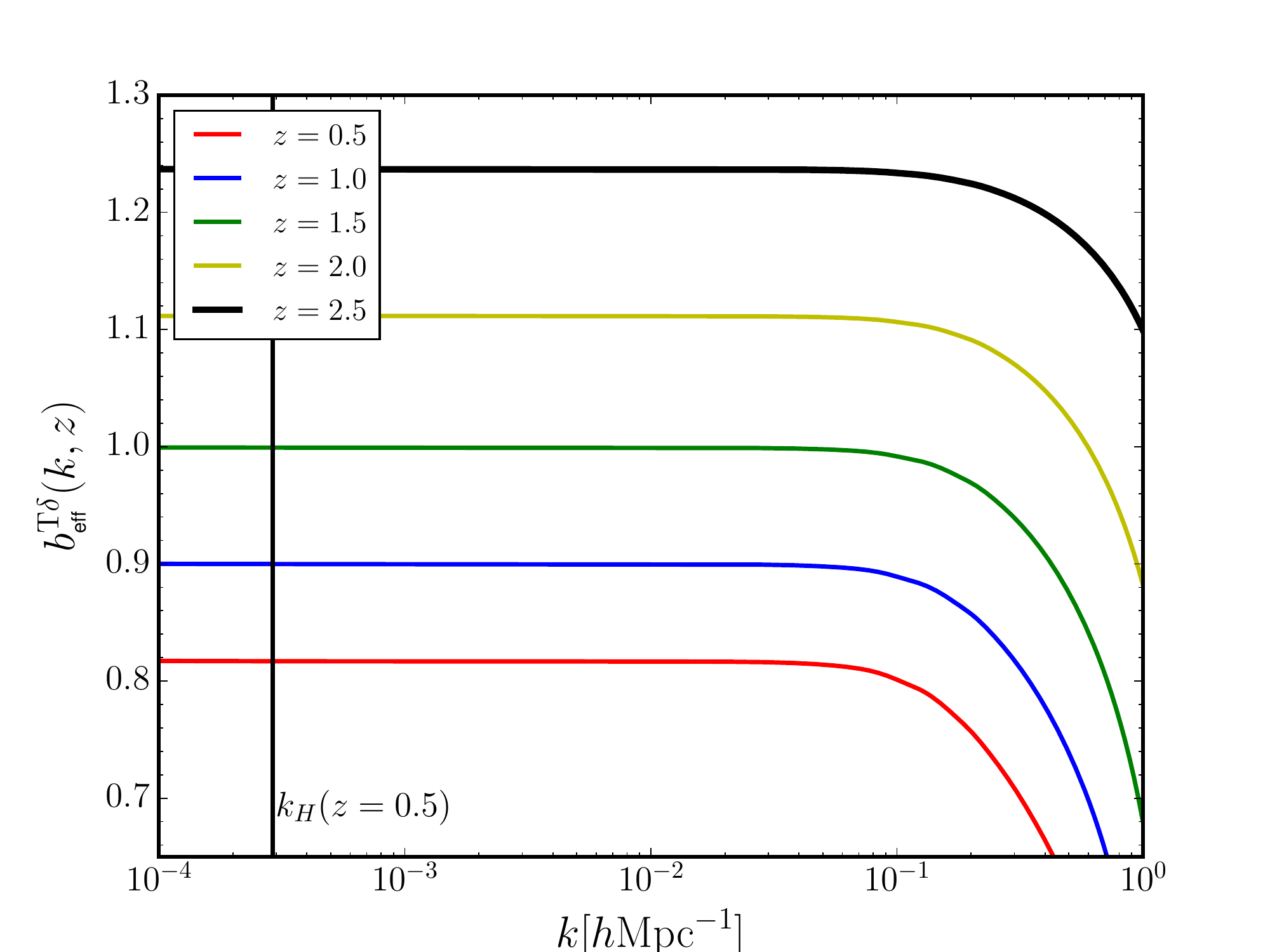}
\caption{\emph{Left panel}: Real space  HI power spectrum with the stochastic power spectrum  contribution subtracted off at five different redshifts $z = [0.5,1.0,1.5,2.0,2.5]$. In the lower left panel we show the fractional difference or the error associated with using the linear power spectrum instead of the full HI power spectrum. \emph{Right panel:} Effective HI bias parameter defined in equation \eqref{eq:effectivebias}. We have normalized $b_{\rm{eff}}^{T\delta}$ to remove dependence on $\bar{T}$. The dark vertical line indicates the horizon scale at $z =0.5$.
 }
\label{fig:PkHI}
\end{center}
\end{figure*}

We show in the right panel of figure \ref{fig:PkHI} the effective or renormalized HI bias parameter(equation \ref{eq:effectivebias}) assuming Sheth-Tormen model gives a fair estimate.  In the left panel we show how it re-scales the power spectrum at one-loop order by computing the frame difference: 
\begin{eqnarray}\label{eq:fractionaldiff}
\Delta_{P_T}(k, z) = \frac{P_{T}(k, z) -S_{n}(z) - P^{11}_{T}(k, z)}{P^{11}_{T}(k, z)}\,.
\end{eqnarray}
There are two important consequences arising from equation \eqref{eq:discretepk} for $T^\obs$ when nonlinear corrections are included in the analysis. These consequences are:
\begin{enumerate}
\item On large scales, nonlinear effects lead to an effective bias parameter which is different from the linear bias parameter estimated from the halo or HI mass function:
\begin{eqnarray}
b_1 \mapsto b_{\text{eff}}^{T \delta} = b_1 + \frac{1}{2}\left(  b_3 + \frac{68}{21}b_2 \right) \sigma^2_{k\rm{nl}}\,,
\end{eqnarray}
The terms in the bracket on the RHS are nonlinear bias parameters, multiplied by the matter variance. When the bias parameter is estimated from observation using the two-point correlation function, the measured value of the bias parameter will correspond to $b_{\text{eff}}^{T \delta}$ and not $b_1$.

\item Secondly, nonlinear effect induces a stochastic power spectrum which becomes important on large scales. The existence of the stochastic power spectrum indicates the break-down of perfect traceability by HI of the underlying matter density field. 
\end{enumerate}
Finally, in addition to $ S_n^{\rm eff}$, the total power spectrum of the HI brightness temperature is also contaminated by the instrumental noise, galactic and extra-galactic foreground. However, most of the analysis of total power spectrum have so far neglected the contribution from  $S_n^{\rm eff}$\cite{Bull:2014rha}. The motivation for this is that the effect of $P_{T}^{\text{shot}}$ for the intensity mapping experiment is negligible\cite{Bull:2014rha,Santos:2015gra,Battye:2016qhf}.  It would be interesting to see whether the smallness of $P_{T}^{\text{shot}}$ also implies  that $ S_n^{\rm eff}$ is negligible. Following \cite{Bull:2014rha}, $P_{T}^{\text{shot}}$ may be calculated from the comoving number density of haloes 
\begin{eqnarray}
P_{T}^{\text{shot}} = \left(\frac{\bar{T}(z)}{\rho_{\HI}(z)}\right)^2\int_{M_{\rm{min}}}^{M_{\rm{max}}} \d M \left[ M_{\HI}^2(M)
n_h(z,M)\right]\,,
\end{eqnarray}
where $\rho_{\HI}(z)$, $n_h$, $M_{\HI}$ and the integral limits are defined in the appendix.

\begin{figure*}[h!]
\begin{center}%
\includegraphics[width=85mm,height=70mm]{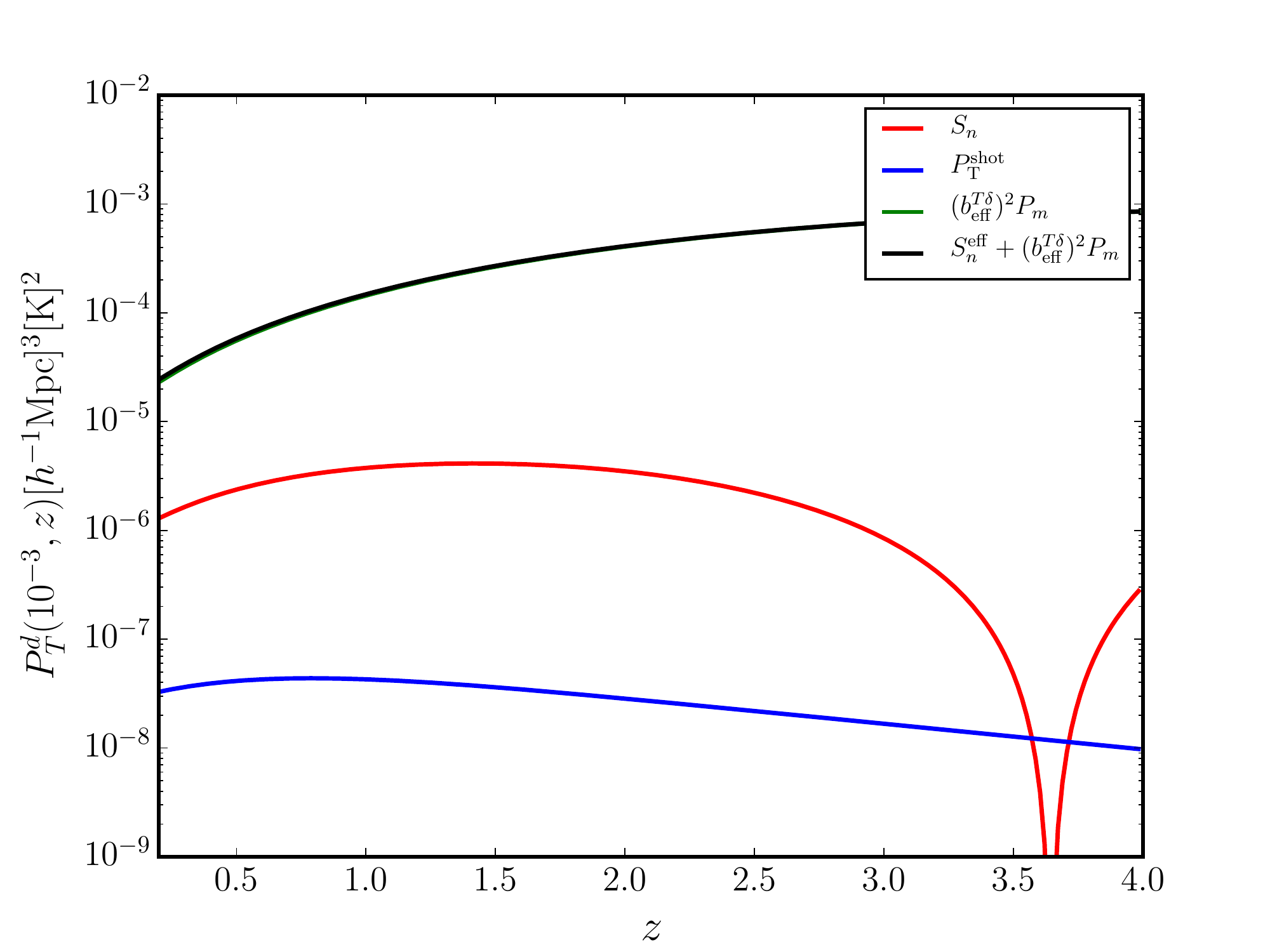}
\includegraphics[width=85mm,height=70mm]{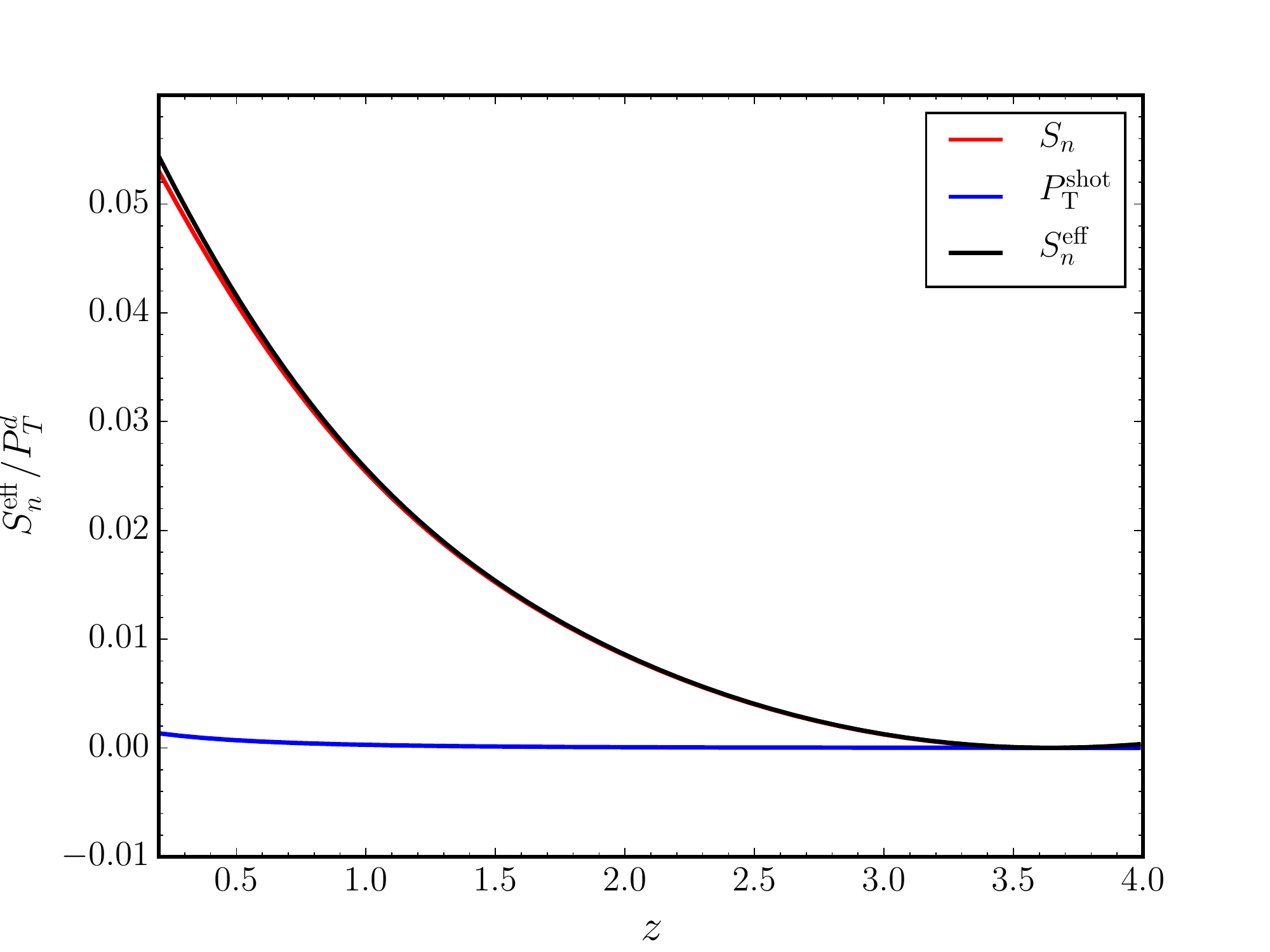}
\caption{\emph{Left panel:} Redshift evolution of the stochastic power spectrum, shot noise, the continuous power spectrum and discrete power spectrum. We set $k = 10^{-3} h \rm{Mpc}^{-1}$. \emph{Right panel:} The fractional contribution of the stochastic power spectrum and  shot noise to the discrete power spectrum with $k$ set at $ 10^{-3} h \rm{Mpc}^{-1}$. 
 }
\label{fig:shotnoise}
\end{center}
\end{figure*}
The renormalization procedure we have described provides a framework for interpreting the stochastic power spectrum as part of the total noise budget. We show in figure \ref{fig:shotnoise} that the contribution from $S_n$ is greater than that from $P_{T}^{\text{shot}}$ for the HI intensity mapping experiment.   And that the effective shot noise contribution is dominated by $S_n$. The left panel shows that $S_n$ constitutes about 5\% of discrete power spectrum at $z \le 0.5$ and $k = 10^{-3} h \rm{Mpc}^{-1}$. This is likely to have implications for the Fisher forecast analysis since the derivative of $S_n$ wrt the cosmological parameters is non-vanishing. Note that $S_n$ does not depend on any scale chosen to regulate the integral.

\subsection{HI power spectrum in redshift space}

\begin{figure*}[h!]
\begin{center}%
\includegraphics[width=85mm,height=70mm]{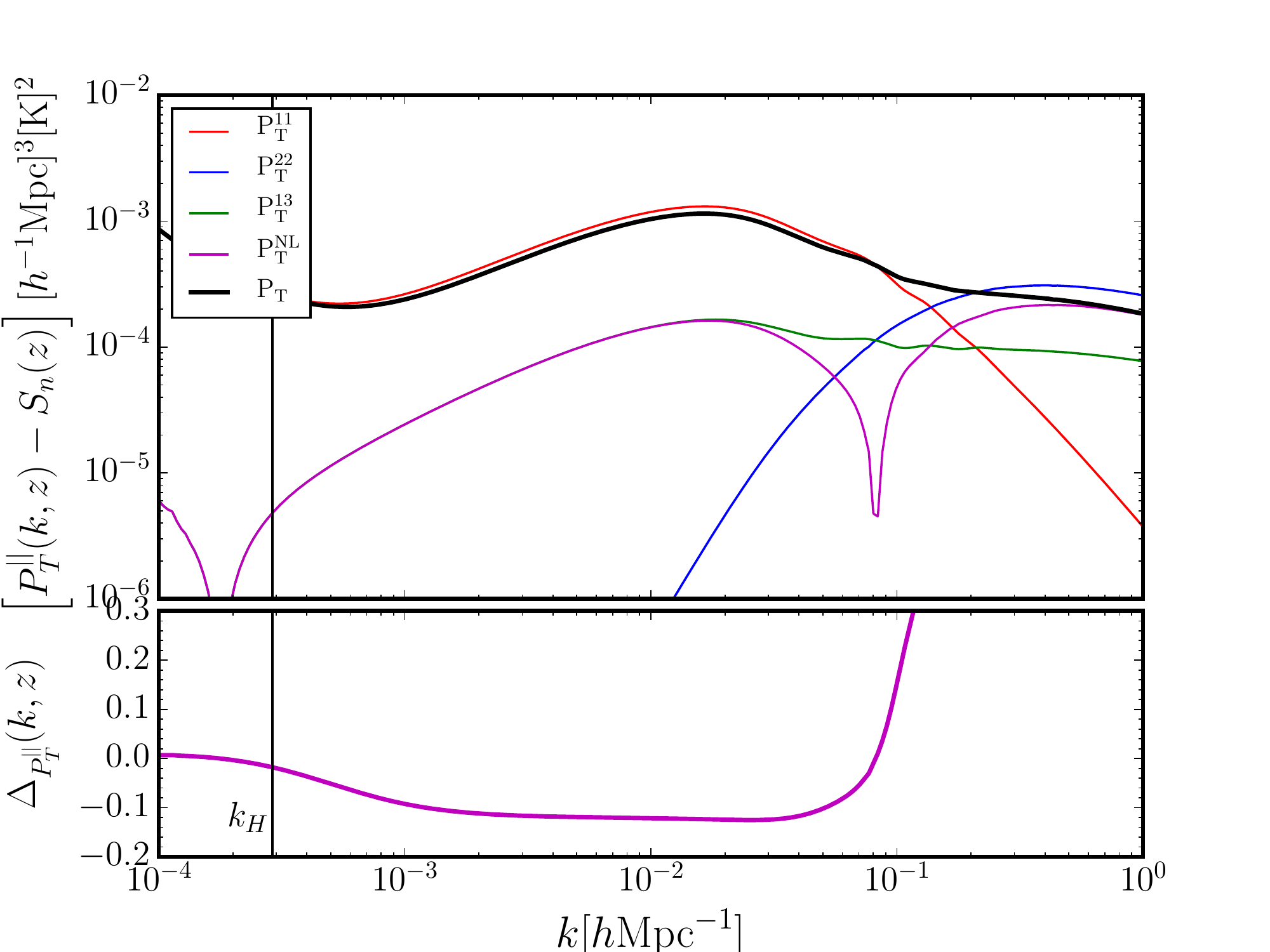}
\includegraphics[width=85mm,height=70mm]{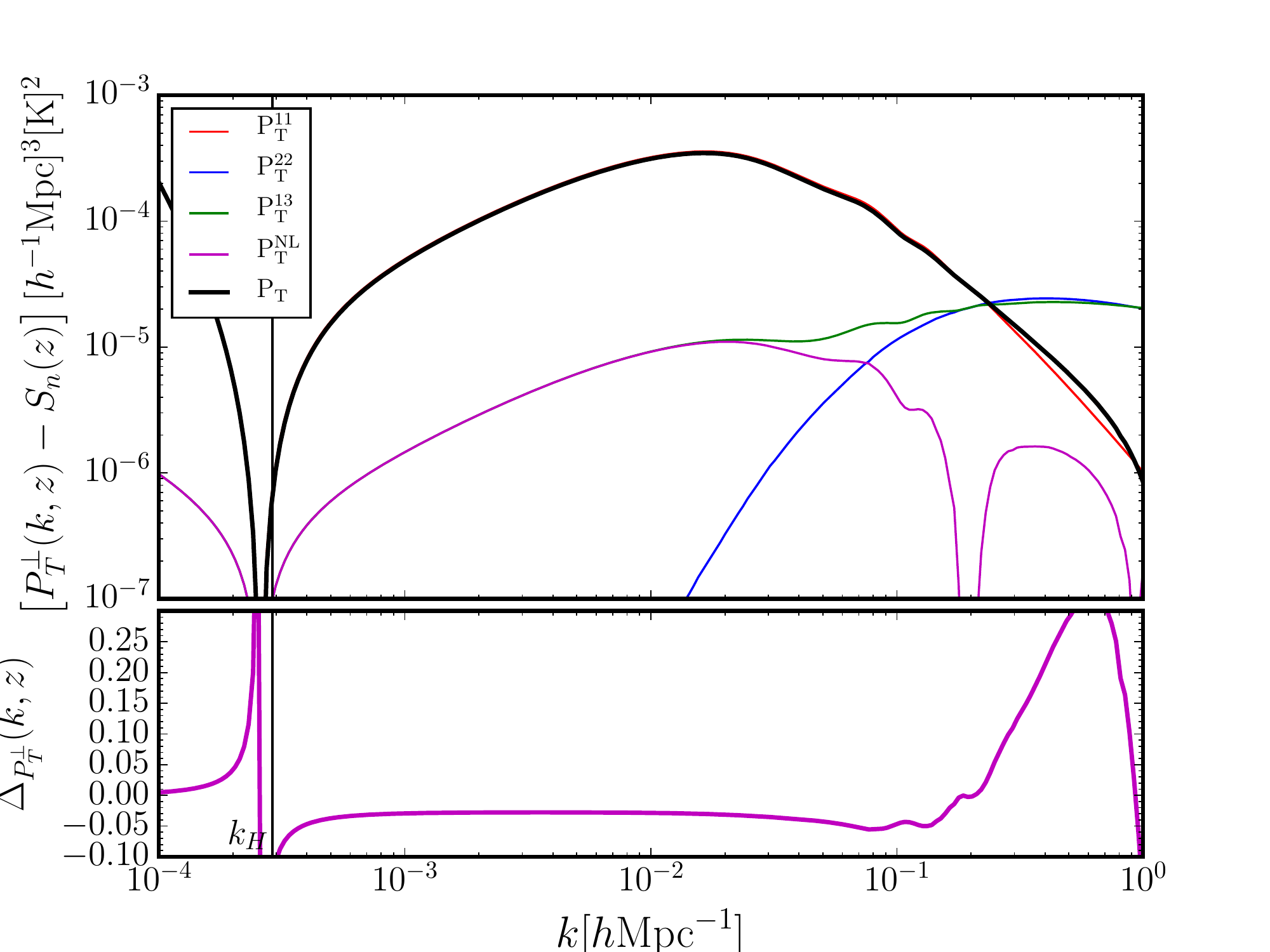}
\caption{ We show the radial ($P_{ T}(k_{\parallel},z) = P^{||}_{ T}(k,z)$) and transverse ($P_{ T}(k_{\bot},z) =P^{\bot}_{ T}(k,z)$ ) power spectrum  at $z = 0.5$ in the left and right panel respectively. The linear part of the transverse power spectrum goes to zero at about $k \sim k_{\HH}$, this feature was first reported in \cite{Yoo:2010ni} for the galaxy power spectrum. In the lower sections of each panel,  we show the fractional difference between the total  power spectrum and the linear theory  prediction.}
\label{fig:RSDPk}
\end{center}
\end{figure*}

For the HI power spectrum in redshift space, we do not assume any phenomenological model for the FoG effect since our key interest is on the ultra-large scale features of the power spectrum.
We show in figure \ref{fig:RSDPk} the relative  contributions of ${P}_{T}^{11}$, ${P}_{T}^{22}$ and ${P}_{T}^{13}$  to the total radial power spectrum(top left panel) and  transverse power spectrum(top right panel). 
The plots show that in addition to the well-known imprint of nonlinear effects on the power spectrum on sub-horizon scales, 
 the one-loop term  gives  a non-vanishing contribution on ultra-large scales, i.e $ k \le k_{\rm{eq}}$ through what may be considered as an additional correction to the bias parameter:
\begin{eqnarray}
 b_{\rm{eff}}^{T\delta}(k,\mu) &\approx & b_1 + \frac{1}{2}\left[\left(  b_3 + \frac{68}{21}b_2 \right) \sigma^2_\Lambda+ \mathcal{I}_{R}(k,\mu)\right]\,.
 \label{eq:RSbias2}
 \end{eqnarray}
The additional term, $\mathcal{I}_{R}$ is due to nonlinear redshift space distortions(see equation \eqref{eq:PT13}). For $\mu=0$, we recover the real space effective HI bias parameter(see equation \eqref{eq:effectivebias}).  The fractional difference 
is weakly scale dependent in redshift space, this is due to the contribution from $\mathcal{I}_{R}$.
  This ultra-large scale dependent imprint of nonlinear effects is absent in the phenomenological models of the power spectrum in redshift space \cite{Scoccimarro:2004tg}, since they assume a simple linear map of  the galaxy over-density to the matter over-density\cite{Taruya:2010mx}.
 
In the lower panels of figure \ref{fig:RSDPk}, we quantify the effective  contribution of  nonlinear effects to the power spectrum explicitly by computing the fractional difference.  We show  in the lower left panel of figure \ref{fig:RSDPk}, that the error could be up to 12\% on equality scale $k_{\rm{eq}}$. 
In the lower right panel of the same figure, we find about 5\% correction for the orthogonal power spectrum, at $k_{\rm{eq}}$. The linear $P_T$ goes to zero on horizon scale, $k_{\HH}$, this is due to vanishing of RSD and Doppler  correction in the transverse direction. Therefore, the effective  contribution is due to gravitational redshift term and $\delta_{\HI}$ and the contribution from their cross correlation  is negative in the neighbourhood of the horizon scales. 
In redshift space, the effective $P_T$ on large scales may be approximated by
\begin{eqnarray}\label{eq:effectiveterm}
P_T^{L}(k) &\approx & \bar{T}^2\bigg\{\left(b_{\rm{eff}}^{T\delta}(k,\mu) +f\mu^2 +\mathcal{A}\,\frac{\HH^2}{k^2}\right)^2 +  \mu^2 \mathcal{B}^2\frac{\HH^2}{k^2}\bigg\}P_{m}({k}) 
+ S_n \,.
\end{eqnarray}
Equation \eqref{eq:effectiveterm} reduces to equation \eqref{eq:largescalePk} in the limit where all the light-cone projection effects  are set to zero. 
 On small  scales, the product of two linear order RSD terms (Kaiser term) contained in $P^{22}_T$ is the dominant term:
 \begin{eqnarray}\label{eq:leadingorderp22}
  {P}_{T}^{22}({k},\mu) &\approx &\frac{\bar{T}^2}{2\pi^2}f^2\int_{k_{\rm{min}}}^{k_{\rm{max}}} {\d k_1}\int_{-1}^{1}\d\mu_k \bigg[ k_1^2\bigg|
 \frac{\mu_1^2(k \mu - k_1\mu_1)^2}{|{\k}-{\k}_1|^2}\bigg|^2
P_{m}(|{\k}-{\k}_1|)P_{m}({k}_1) \bigg]\,.
 \end{eqnarray}
 Equation \eqref{eq:leadingorderp22} vanishes  for $\mu =0$. 
We conclude this section by computing the monopole of the HI power spectrum 
\begin{eqnarray}\label{eq:monopolepower}
P^{0}_T(k)&=& (\bar{T})^2\bigg[b_1^2 +\frac{2}{3} b_1  f+ \frac{1}{5}f^2+ \frac{1}{3} \left[\mathcal{B}^2
+ 2\left( 3b _1 + f \right) \mathcal{A}\right]\frac{\HH^2}{k^2}
+\mathcal{A}^2\frac{\HH^4}{k^4}\bigg]P_m(k) 
 +\frac{1}{2} \int_{-1}^{1}\d \mu\,\left[  {P}_{T}^{\rm{NL}}({k},\mu)\right]\,.
\end{eqnarray}
\begin{figure*}[hbt!]
\begin{center}%
\includegraphics[width=80mm,height=70mm]{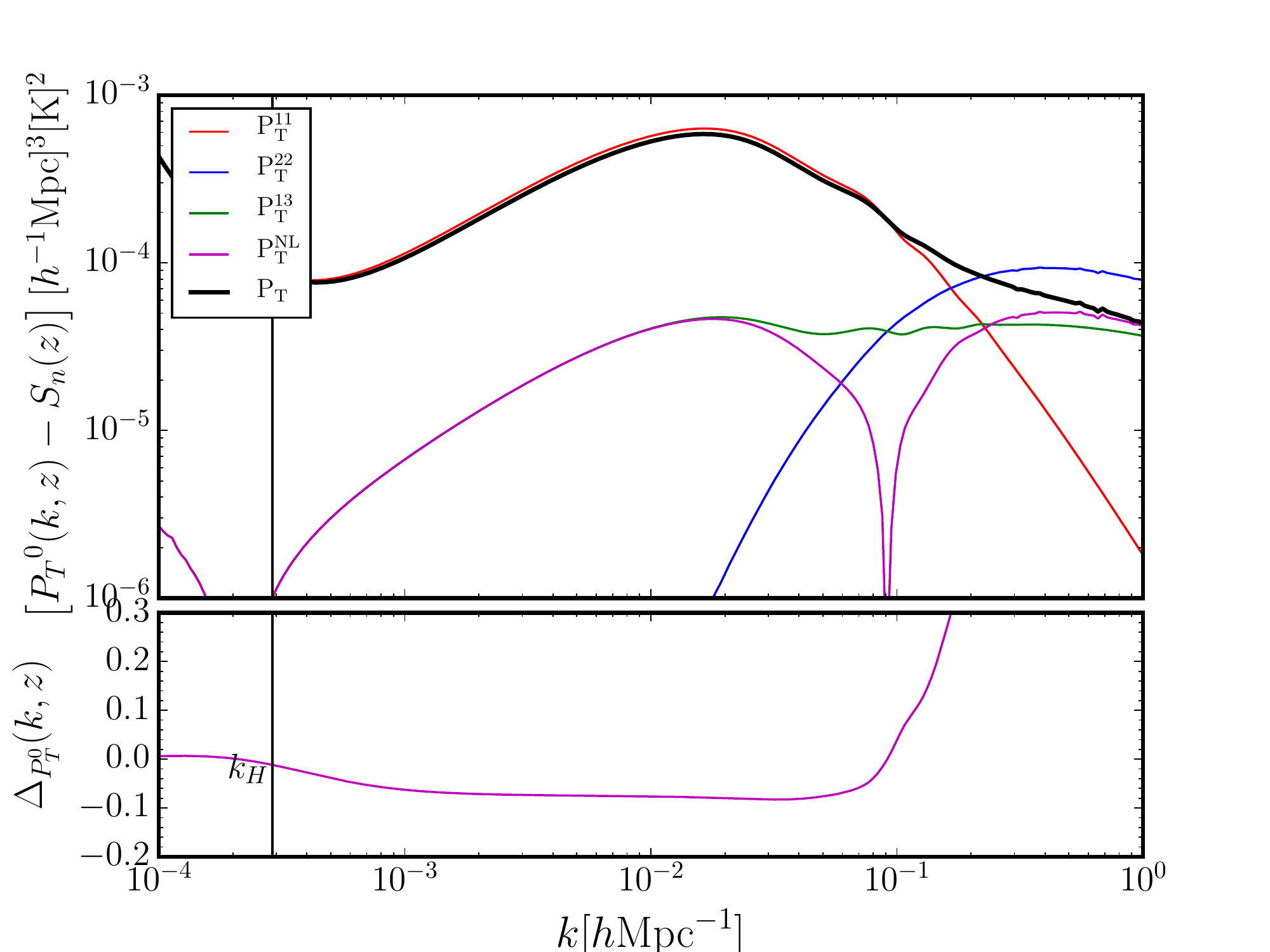}
\includegraphics[width=80mm,height=70mm]{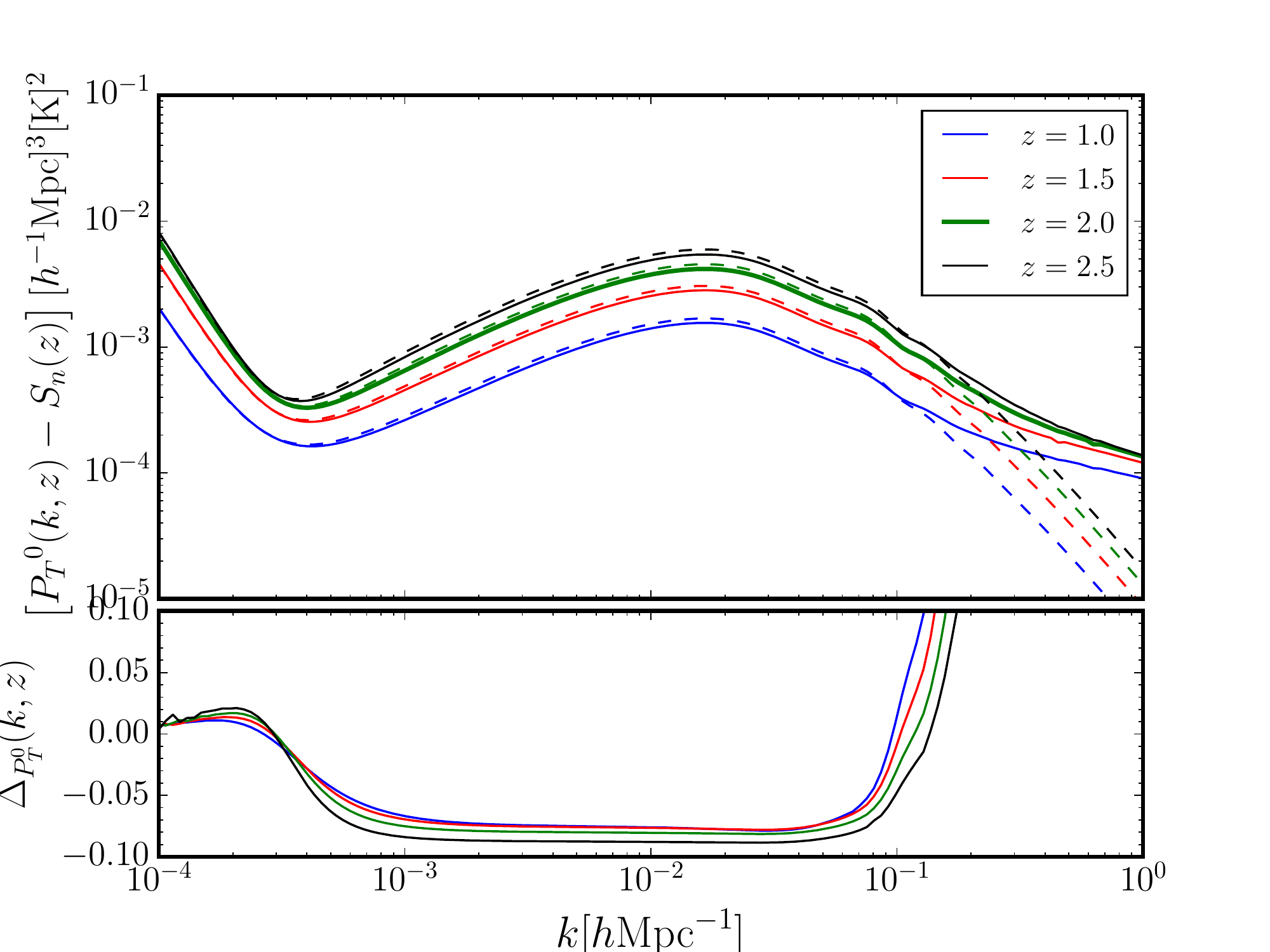}
\caption{In the left panel, we show the  monopole of the total  power spectrum of the HI brightness temperature in  redshift space at $z =0.5$. One the  right panel, we show the monopole of the power spectrum of $T\obs$ at $z =1.0$ (blue curve), $z =1.5$ (red curve), $z =2.0$ (green curve) and $z =2.5$ (black curve). The dashed lines indicate the linear theory prediction. The fractional difference is shown in the bottom panels.}
\label{fig:pkmonpole}
\end{center}
\end{figure*}

At $z =0.5$, there is about 7\% modulation in the amplitude of the monopole of $P_T$ by nonlinear effects  on large scales. The effective contribution drops to less than 2\% at the horizon scale. In the right panel of figure \ref{fig:pkmonpole}, we show the redshift dependence of nonlinear effects on the power spectrum. The magnitude of the nonlinear correction  increases slightly as the  redshift increases. This feature is due to the redshift evolution of the bias parameters. We summarise the  imprints of nonlinear effects in the bias and nonlinear RSD  on$P_T$ on all scales  in Table \ref{tab:one}. We did not calculate the effective HI bias in redshift space in a similar way as we did in real space, this will be pursued in a future work since it will require the monopole of the matter power spectrum in redshift space which is beyond the scope of this work.
\begin{table}[!h]
\centering
\begin{tabular}{|c|c|c|c|}
\hline
$P_{T}(k)$ & $k_{\HH} \le k \le k_{\rm{eq}}$ &$k_{\rm{eq}} \le k \le k_{\rm{BAO}}^{\rm{end}}$& $k > k_{\rm{BAO}}^{\rm{end}}$  \\ \hline
$P_{T}^{\|}(k)$ & $1\,\le$\,$ \Delta \le\,$ 12 &12\,$\le$\,$ \Delta \le\, 30$ & $ \Delta > 30$   \\ \hline  
$P_{T}^{\bot}(k)$& $10$\,$\le$\,$ \Delta  \le\,$5 & 5\,$\le$\,$ \Delta  \le\,$5 & $\Delta > 10$   \\ \hline
$P_{T}^0(k)$ & $1\,\le$\,$ \Delta \le\, 7$ & $7 \le$\,$ \Delta \le\,30$ & $\Delta >30$   \\ \hline
\end{tabular}
\caption{Summary of the imprint of nonlinear effects in the bias and nonlinear RSD on HI power spectrum in redshift space at $z =  0.5$. $\Delta  $ corresponds to the percentage error associated with using linear theory approximation to calculate the  HI power spectrum. We  defined $k_{\rm{BAO}}^{\rm{end}} \approx 0.3 \,h\rm{Mpc}^{-1}$. Predictions of the standard perturbation theory at one-loop level may not be trusted beyond $k_{\rm{BAO}}$ at $z =0.5$.}
\label{tab:one}
\end{table}

\section{Conclusion}\label{sec:conc}

In this paper, we have studied in detail, how nonlinearity in the bias and peculiar velocity could modulate clustering of HI brightness temperature on large scales.  We approached the problem  by using the general relativistic perturbation theory techniques to derive for the first time, the full nonlinear expression for the  HI brightness temperature up to third order in perturbation theory in the post-re-ionization universe.  The full expression  we derived,  describes how weak gravitational lensing, redshift space distortions induced by the peculiar velocity, gravitational redshift, ISW and HI bias  affect  the clustering of the HI brightness temperature.  When we take the plane-parallel approximation, we recover exactly the previous results that made use of a different technique to derive the corresponding expression for the galaxy number count in the Newtonian limit\cite{Bernardeau:2001qr,Heavens:1998es}. 

Our expression for the perturbed HI brightness temperature includes contributions from: the Sachs-Wolfe effect(gravitational redshift correction), Integrated Sachs-Wolfe effects(contribution due to evolving gravitational potential along the line of sight), Doppler effects(effect due to relative motion of the source and the observer), Kaiser effect(linear RSD), nonlinear RSD (FoG effect) and weak gravitational lensing. We introduced the HI bias after transforming the HI over-density to the comoving synchronous gauge, then use simple local Eulerian bias model expanded up to third order in perturbation theory to relate it to the underlying dark matter density field. The HI bias parameters were calculated from the halo model using Sheth-Torman halo mass function\cite{Sheth:2001dp}. We allowed non-Gaussianity in the bias only in the  linear order perturbation of HI number density, since its influence on the bias is sub-dominant at one-loop level. The signal from non-Gaussianity in the matter density field  is also sub-dominant at power spectrum level, hence we neglected it as well. 

We calculate the real space power spectrum of the HI brightness temperature at four different redshift points. We use a bias renormalization prescription presented in \cite{McDonald:2009dh} to renormalize FLRW background number density of HI atoms and the brightness temperature. The auto power spectrum of the HI bias parameter at second order leads to a white noise-like contribution on large scales, this contribution may be treated as part of the effective shot noise contribution.  
We also calculate the redshift space power spectrum of the HI brightness temperature. We show that, in addition to nonlinear effects contributing significantly to the  power spectrum on small scales, it also modulates the  power spectrum  on large scales due to mode coupling at nonlinear order. We show that on scales  $k \le k_{\rm{eq}}$, $P_T$ receives a cut-off dependent correction from nonlinear effects in the bias and nonlinear RSD. 

Finally, all the results we presented on modulation of the HI power spectrum on large scales by nonlinear effects is dependent on the cut-off chosen to regulate the convolution integrals in $P^{13}_{T}$. We adopted a simple regularisation technique which suppresses modes greater than $k_{\rm{nl}}$. This cut-off is chosen to ensure that we remain within the regime of validity of  the cosmological  perturbation theory\cite{Smith:2002dz}. A more rigorous treatment  will involve adopting a renormalization procedure that could eliminate the dependence of the result on a cut-off scale\cite{Baldauf:2014qfa}. Also, we used a simple Sheth-Tormen halo mass function\cite{Sheth:1999mn} and a polynomial fit for the HI mass profile to estimate all the HI bias parameters. A more accurate model for the HI mass would be needed for a more accurate quantitative treatment\cite{Padmanabhan:2016fgy,Padmanabhan:2016odj}.

\vspace{8mm}

\acknowledgments  
I would like to  thank  Roy Maartens and Mario Santos who were involved in the early stages of this project.  I would also like to thank 	Aur\'elie P\'enin, Jose Carlos S Fonseca, and  Ravi Sheth for extensive discussions. I thank Bishop Mongwane   for comments on the draft.
I am supported by the South African Square Kilometre Array Project.  The algebraic computation in the paper was done with the help of the tensor algebraic perturbation theory software xPand \cite{Pitrou:2013hga}.

\appendix

\section{Basic cosmological  perturbation theory}\label{sec:LOS}

Here we provide more details on the perturbation theory calculation up to third order, using the approach developed in \cite{Umeh:2012pn,Umeh:2014ana}.
We consider a perturbed FLRW spacetime in the Poisson gauge assuming  a conformally flat background metric
\begin{eqnarray}\label{eq:metric}
\d s^2=a^2\left[-(1 + 2\Phi+\Phi\two + \frac{1}{3}\Phi\three )\d \eta^2 
 + \left((1-2 \Psi-\Psi\two -\frac{1}{3}\Psi\three)\delta_{i j} \right)\d x^{i}\d x^{j}\right]\,.
\end{eqnarray}
Here the  proper time is related to conformal time  by   $\d t= a\d \eta$.  $\Phi$ is the first-order Newtonian gravitational potential, $\Psi$ is scalar curvature perturbation. 
For any  tensor $S$ we  expand it up to third order as
\begin{equation}\label{eq:stdprescription}
 \hat{S}=\bar{S}+\delta\one S +\frac{1}{2}\delta\two S + \frac{1}{3!} \delta \three S\,.
\end{equation}
We expand the 4-velocity $u^a$,  of a matter field using  (\ref{eq:metric}) up to third order
\begin{eqnarray}
u^0&=&1 - \Phi +  \frac{1}{2}\left[3 \Phi^2 - \Phi\two + {\D}_{i}v{\D}^{i}v\right] + \frac{1}{3!}\bigg[-\Phi\three - 15 \Phi^3 +  9 \Phi \Phi\two + 3 {v\two}^i{\D}_i v - 3 \Phi {\D}_i v {\D}^iv 
\\ \nonumber &&
- 6\Psi{\D}_i v {\D}^i v + 3 {\D}_iv\two {\D}^i v\bigg] \,,\\
u^i&=&{\D}^{i}v + \frac{1}{2}
\left[v^{i}{}\two +  {\D}^{i}v\two\right] + \frac{1}{3!}
\left[v^{i}{}\three +  {\D}^{i}v\three\right]\,.
\end{eqnarray}

In cosmological perturbation theory, the map between real and redshift space is given by $s^i = x^i + \Delta\one x^i + \Delta\two x^i/2 + \Delta\three x^i/3!$, where
\begin{eqnarray}\label{eq:map1}
\Delta\one x^i &=&\overbrace{ \delta\one x^i_{\bot}}^{\Delta\one x_{\bot}} + n^i\overbrace{\left(\delta\one x\p
+ \frac{\delta\one z}{\HH_s}\right)}^{\Delta\one x\p}
\approx  \delta\one x^i_{\bot} + n^i \frac{\delta\one z}{\HH_s}\,.
\end{eqnarray}
Here, we have considered only the dominant terms in the second equality. In the plane-parallel limit the dominant term is given by 
\begin{eqnarray}
\Delta\one x^i &\approx &n^i \frac{\delta\one z}{\HH_s} = -n^i \frac{1}{\HH_s}\nabla_{\p}v_{s}\one \,.
\end{eqnarray}
We have considered only the radial component of the peculiar velocity.  The second order correction to the real to redshift space map is given by
\begin{eqnarray}
\Delta\two x^i &=&\delta\two x^i_{\bot} 
 -\frac{2}{\HH_s}\delta\one x'_{\bot}{}^i \delta\one z+2\Delta\one x\p\nabla\p\delta\one x^i_{\bot} + 2 \Delta\one x^i_{\bot} \nabla_{\bot j}\delta\one x^i_{\bot}
\\ \nonumber &&
+ n^i\bigg[\delta\two x\p
+ \frac{\delta\two z}{\HH_s}-2\frac{\delta\one z}{\HH_s}\left(\delta\one x\p' +\frac{\delta\one z'}{\HH_s}\right)+ 2\Delta\one x\p\left(\nabla\p\delta\one x\p + \frac{\nabla\p \delta\one z }{\HH}\right)
\\ \nonumber&&
+2\Delta\one x^j_{\bot} \left(\nabla_{\bot j}\delta\one x\p
 +\frac{\nabla_{\bot j} \delta\one z}{\HH_s}\right) -\frac{2}{\chi} \delta\one x^k_{\bot} \Delta\one x_{\bot k}
-\frac{(\delta\one z)^2}{\HH_s}\left(1+\frac{\HH'_s}{\HH_s^2}\right)\bigg]
\\ 
&\approx& \overbrace{\delta\two x^i_{\bot} + 2 \Delta\one x^j_{\bot} \nabla_{\bot j}\delta\one x^i_{\bot}}^{\Delta\two x_{\bot}}+\frac{n^i}{\HH_s} \overbrace{\left[\delta\two z  +2\left(\Delta\one x\p \nabla\p \delta\one z +\Delta\one x^j_{\bot}\nabla_{\bot j} \delta\one z\right)\right]}^{\Delta\two x\p}\,,
\end{eqnarray} 
where the approximation sign  shows the limit of  approximation used in the paper.  The radial component becomes 
\begin{eqnarray}
\Delta\two x\p &\approx &\frac{n^i}{\HH_s} \left[\delta\two z  +2\left(\Delta\one x\p \nabla\p \delta\one z +\Delta\one x^j_{\bot}\nabla_{\bot j} \delta\one z\right)\right]
\,,
\\
&\approx &n^i\left[-\frac{1}{\HH_s}\nabla_{\p}v_{s}\two  + 2\left[\frac{1}{\HH_s^2}\nabla_{\p}v_{s}\one\nabla_{\p}^2v_{s}\one - \frac{1}{\HH_s}\nabla_{\bot j}\nabla_{\p}v_{s}\int^{\chi_s}_{0}\frac{(\chi-\chi_s)}{\chi_s}\nabla_{\bot}^j\Phi_A\one\d\chi\right]\right]\,,
\end{eqnarray}
The terms in the square bracket in the second equality are  Post-Born terms.
 The covariant derivatives which appear after implementing Born and Post-Born correction to the fluctuation of the HI brightness temperature are decomposed as follows:
\begin{eqnarray}
\Delta\one x^a\nabla_{a}\delta \one T &=& \frac{\delta\one z}{\HH_s} {\delta\one{T}}' + \Delta\one x_{\p} \nabla_{\p} \delta\one{T} + \Delta\one x_{\bot}^i\nabla_{\bot i} \delta\one{T} \,, \\
&\approx& \Delta\one x_{\p} \nabla_{\p} \delta\one{T} + \Delta\one x_{\bot}^i\nabla_{\bot i} \delta\one{T} \,,
 \end{eqnarray}
 where $\Delta\one x_{\p}$ and $\Delta\one x_{\bot}^i$ are given in equation \eqref{eq:map1}. For double covariant derivatives we decompose as follows:
 \begin{eqnarray}
 \Delta\one x^a\Delta\one x^b \nabla_{a}\nabla_{b} \delta \one T &=&
(\delta\one \lambda)^2 {\delta \one T }'' - 2 \delta\one\lambda \Delta\one x\p \nabla\p{\delta \one T  }' + (\Delta\one x\p)^2 {\nabla\p}^2{\delta \one T  }
\\ \nonumber &&
+ \frac{1}{\chi_s} \Delta\one {x_{\bot}}_i\Delta\one {x_{\bot}}^i \nabla\p {\delta \one T  }
-\frac{2}{\chi_s}\Delta\one x\p\Delta\one {x_{\bot}}^i\nabla_{\bot i} \delta \one T 
+ 2\Delta\one x\p\Delta\one {x_{\bot}}^i\nabla_{\bot i}\nabla\p \delta \one T 
\\ \nonumber &&
+ 2 \delta\one \lambda \Delta\one {x_{\bot}}^i\nabla_{\bot i} {\delta \one T }'
+ \Delta\one {x_{\bot}}^i\Delta\one {x_{\bot}}^j\nabla_{\bot i} \nabla_{\bot j}{\delta \one T }
\,,
\\  
  &\approx & (\Delta\one x\p)^2 \nabla\p^2\delta \one T  +2\Delta\one x\p\Delta\one {x_{\bot}}^i\nabla_{\bot i}\nabla\p \delta \one T  +\Delta\one {x_{\bot}}^i\Delta\one {x_{\bot}}^j\nabla_{\bot i} \nabla_{\bot j}{\delta \one T }\,,
\end{eqnarray}
where in the second equality, we have considered only the dominant part.
We have twice projected screen space derivatives, we expand them in Fourier space by first expressing them in terms of the spatial derivatives
\begin{eqnarray}
\nabla_{\bot}^2 &=& \nabla_i\nabla^i - \nabla\p^2 -\frac{2}{\chi}\nabla\p \approx \nabla_i\nabla^i - \nabla\p^2\,, \\
 \nabla_{\bot i}\nabla_{\bot j} &=& \nabla_{i}\nabla_{j}
 -n_in_j \nabla^2\p- 2n_{(i}\nabla\p \nabla_{\bot j)} -\frac{1}{2\chi} N_{ij} \nabla\p  + \frac{1}{\chi} n_{(i} \nabla_{\bot j)} \approx  \nabla_{i}\nabla_{j}
 -n_in_j \nabla^2\p- 2n_{(i}\nabla\p \nabla_{\bot j)}\,.
\end{eqnarray}
And for once projected screen space derivative and derivative along the line of sight we have
\begin{eqnarray}
\nabla\p = n^i \nabla_i \,,\quad \nabla^2\p = \nabla\p\nabla\p\,, \quad
\nabla_{\bot i} = N_i{}^j \nabla_j  = \nabla_i - n_i \nabla\p\,,\quad \nabla_{i}n^j = \frac{1}{\chi} N^j{}_i
\end{eqnarray}

\section{Standard Dark matter perturbation theory}\label{sec:darkmatter}

The solutions for the $n$-th order solution for a coupled dark matter over-density $\delta_\text{m}$ and velocity divergence, $\theta({\k})$ equation are given by \cite{Bernardeau:2001qr}: 
\begin{align}\label{eq:Newtonaindef}
\delta_\text{m}({\k})=\sum_{n=1}^\infty \delta_\text{m}^{(n)}({\k}), &&
\theta({\k})=-\mathcal{H} f\sum_{n=1}^\infty \theta^{(n)}({\k})\,.
\end{align}
At $n$-th order we have 
\begin{eqnarray}\label{eq:darkmatter}
\delta_\text{m}^{(n)}({\k})=\int \frac{d^3k_1}{(2\pi)^3}\ldots \int
\frac{d^3k_{n}}{(2\pi)^3} \delta_\text{m}({\k}_1)\ldots \delta_\text{m}({\k}_{n})F_{n}({\k}_1,\ldots,{\q}_{n})\delta^\text{(D)}({\q}_1+\ldots+{\k}_{n}-{\k})\,,\\
\theta^{(n)}({\k})=\int \frac{d^3k_1}{(2\pi)^3}\ldots \int
\frac{d^3k_{n}}{(2\pi)^3} \delta_\text{m}({\k}_1)\ldots \delta_\text{m}({\k}_{n})G_{n}({\k}_1,\ldots,{\k}_{n})\delta^\text{(D)}({\k}_1+\ldots+{\k}_{n}-{\k}),
\end{eqnarray}
where the coupling kernels $F_{n}$ and $G_{n}$ can be obtained using recursion
relations. 
\begin{eqnarray}
{F}_2({\k}_1,{\k}_2)&  = & {5\over 7} + {{\k}_1\cdot {\k}_2\over 2 k_1 k_2}
\left({k_1\over k_2} + {k_2\over k_1}\right) + {2\over 7}\left({{\k}_1\cdot
{\k}_2\over k_1 k_2}\right)^2,\\
{G}_2({\k}_1,{\k}_2)&  = & {3\over 7} + {{\k}_1\cdot {\k}_2\over 2 k_1 k_2}
\left({k_1\over k_2} + {k_2\over k_1}\right) + {4\over 7}\left({{\k}_1\cdot
{\k}_2\over k_1 k_2}\right)^2,\\ 
{F}_3^s({\k}_1,{\k}_2,{\k}_3)& = & F_2({\k}_2,{\k}_3)
\left[{1\over 3} + {1\over 3}{{\k}_1\cdot({\k}_2+{\k}_3)\over ({\k}_2+{\k}_3)^2}
+{4\over 9}{{\k}\cdot{\k}_1\over k_1^2}{{\k}\cdot({\k}_2+{\k}_3)\over
({\k}_2+{\k}_3)^2}\right]\\ \nonumber
&&-{2\over9} {{\k}\cdot{\k}_1\over k_1^2}{{\k}\cdot({\k}_2+{\k}_3)\over
({\k}_2+{\k}_3)^2}{{\k}_3\cdot({\k}_2+{\k}_3)\over
k_3^2} + {1\over 9} {{\k}\cdot{\k}_2\over k_2^2} {{\k}\cdot{\k}_3\over k_3^2}\,,\\ 
{G}_3^{s}({\k}_1,{\k}_2,{\k}_3)& = & 3F_3^s({\k}_1,{\k}_2,{\k}_3) - {{\k}\cdot
{\k}_1\over k_1^2}F_2({\k}_2,{\k}_3)- {{\k}\cdot({\k}_1+{\k}_2)\over 
({\k}_1+{\k}_2)^2}G_2({\k}_1,{\k}_2),
\end{eqnarray}
with ${\k}={\k}_1+{\k}_2+{\k}_3$ in the last two expressions.  The subscript $S$ 
indicates that the expression has been made symmetric w.r.t. any permutation
of the arguments. We normalized the dark matter kernels appropriately to agree with the coefficient of our Taylor expansion. 
The matter power spectrum up to one-loop correction is given by
\begin{eqnarray}
{P}_{\delta\delta}(k)  & = & P_m(k) +P^{22}_{\delta\delta}(k) +  P^{13}_{\delta\delta}(k)\,,
\end{eqnarray}
where $P_m(k)$ is the standard linear dark matter power spectrum and other one-loop correction terms are given by
\begin{eqnarray}\label{eq:defmatterpower}
P^{22}_{\delta\delta}(k)&=&\frac{1}{2}\int \frac{\mathrm{d}^3k_1}{(2\pi)^3} P_m(k_1)P_m(\left|{\k}-{\k}_1\right|)\left|F_2({\k}_1,{\k}-{\k}_1)\right|^2\,, \\
P^{13}_{\delta\delta}(k)
&=&\frac{1}{252} \frac{k^3}{4\pi^2} P_m(k) \int_0^\infty dr~ P_m(kr)
        \left[\frac{12}{r^2} - 158 + 100r^2 - 42r^4
            + \frac{3}{r^3}(r^2-1)^3 (7r^2+2) \ln\left|\frac{1+r}{1-r}\right| \right]\,.
\end{eqnarray}
Similarly, for the peculiar velocity term, we have 
\begin{eqnarray}
{P}_{\theta\theta}(k)  & = & P_m(k) +P^{22}_{\theta\theta}(k) +  P^{13}_{\theta\theta}(k)\,,
\end{eqnarray}
where $P^{22}_{\theta\theta}(k)$ and $P^{13}_{\theta\theta}(k)$  take similar forms as in equation \eqref{eq:defmatterpower}.
\begin{eqnarray}
P^{22}_{\theta\theta}(k)&=&\frac{1}{2}\int \frac{\mathrm{d}^3k_1}{(2\pi)^3} P_m(k_1)P_m^{11}(\left|{\k}-{\k}_1\right|)\left|G_2({\k}_1,{\k}-{\k}_1)\right|^2\,, \\
P^{13}_{\theta\theta}(k) &=&
        \frac{1}{84} \frac{k^3}{4\pi^2} P_m(k) \int_0^\infty dr~ P_m(kr)
        \left[\frac{12}{r^2} - 82 + 4r^2 - 6r^4
            + \frac{3}{r^3}(r^2-1)^3 (r^2+2) \ln\left|\frac{1+r}{1-r}\right| \right]\,.
\end{eqnarray}

The integrands introduced in equation \eqref{eq:curlyIintegral} are \cite{Matsubara:2007wj,Donghui2010}
\begin{eqnarray}
B_{1101}(r) &=& \frac{1}{2}\ , \\
B_{1110} (r) &=& \frac{1}{84}\left[-2\left(9r^4 - 24 r^2 + 19 \right) + \frac{9}{r}\left(r^2 -1\right)^3 \ln\left( \frac{1+ r}{|1-r|}\right)\right] \,,\\
B_{1210}(r) &=& -\frac{1}{3} \,,\\
B_{1200}(r) &=& - \frac{1}{336 r^3}\left[2 \left(-9 r^7 + 33 r^5 + 33 r^3 - 9 r\right) + 9 \left( r^2 -1\right)^4 \ln \left(\frac{1+r}{|1-r|}\right)\right]\,, \\
B_{2220}(r) &=& \frac{1}{336 r^3} \left[2r \left( - 27 r^6 + 63 r^4- 109 r^2 + 9\right) + 9 \left(3 r^2 + 1 \right) \left( r^2 -1\right)^3 \ln \left( \frac{1+r}{|1-r|}\right) \right]\,, \\
B_{2300}(r) &=& - \frac{1}{3}\,.
\end{eqnarray}

\section{How we obtain HI bias from halo bias}\label{sec:bias}

\begin{figure}[htb!]
\includegraphics[width=80mm,height=70mm ]{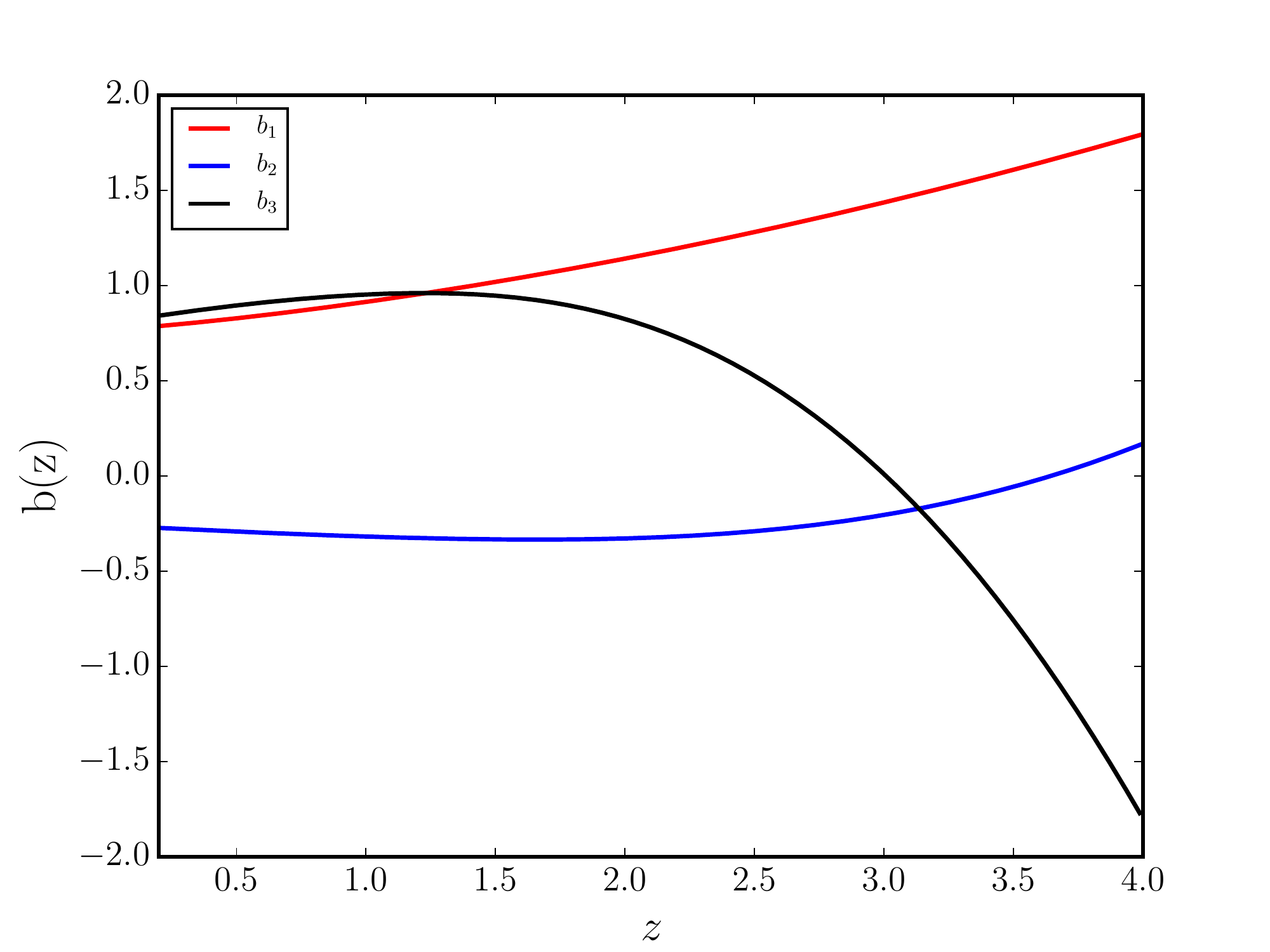}
\includegraphics[width=80mm,height=70mm ]{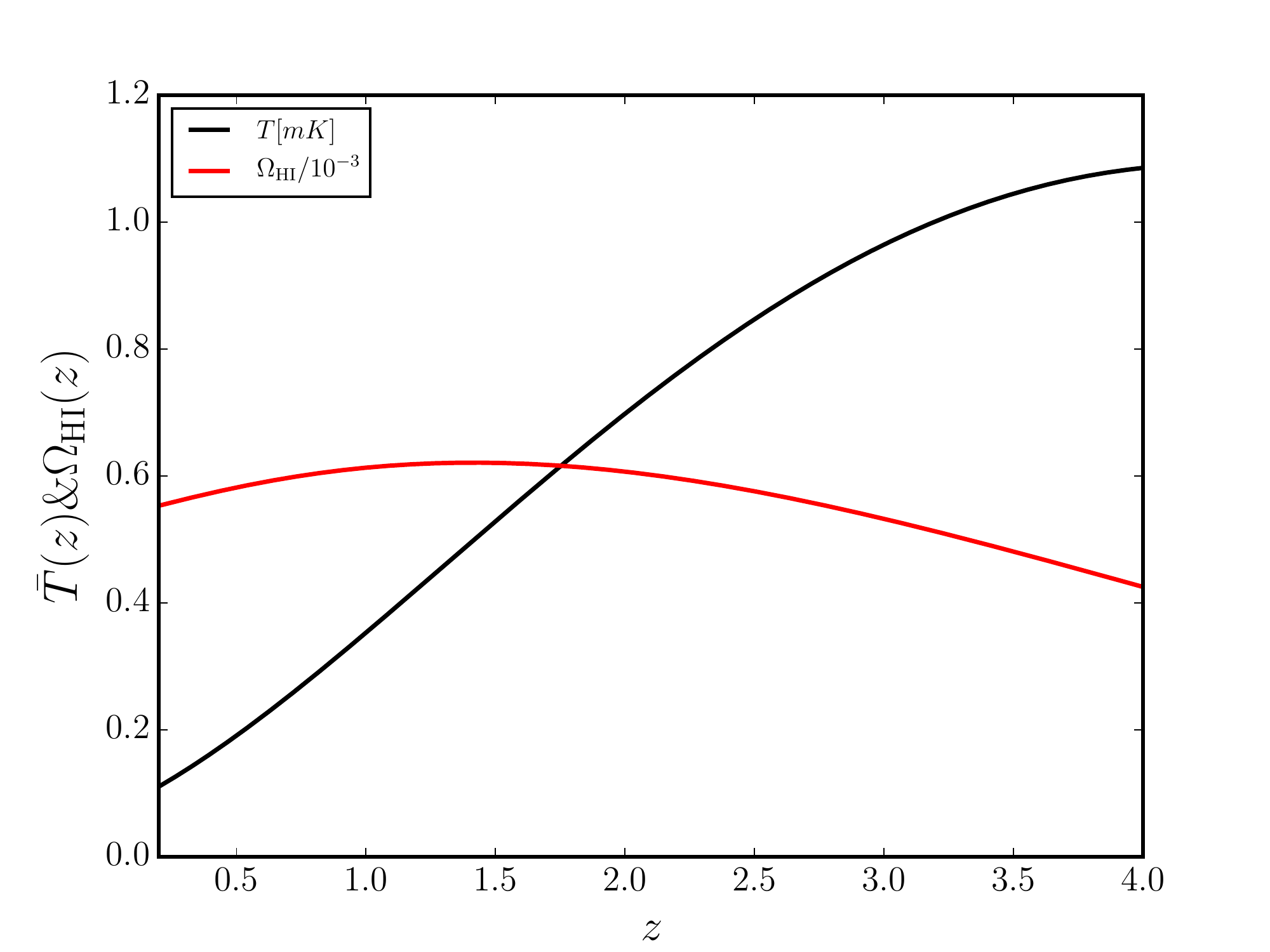}
\caption{\emph{Left panel}: The HI bias parameters constructed from Sheth-Torman halos mass function.  \emph{Right panel}: Redshift evolution of the mean HI brightness temperature  and HI density parameter.}
\label{fig:biasparameters}
\end{figure}

We calculate the bias parameters from a simple Sheth-Torman halos mass function for a spherical collpase model \cite{Sheth:1999mn}: 
\begin{equation}
n_h(M) = \nu f(\nu) \frac{\bar{\rho}}{M^2}\frac{\d\ln \nu}{\d\ln M},
\end{equation}
where the peak height $\nu$ is related to the variance in dark matter density field, $\sigma^2$, $\nu = (\delta_c/\sigma_{nG})^2$ and $\delta_c = 1.686$ is  the critical threshold for a spherical collapse at the current epoch  obtained from linear perturbation theory.  A halo of  mass $M = \bar{\rho} V$  is formed when the walk first crosses a barrier $f(\nu)$:
\begin{equation}
\nu f(\nu)=A(p)\left(1+\frac{1}{(q
\nu)^p}\right)\sqrt{\frac{q\nu}{2\pi}}\left[-\frac{q\nu}{2}\right],
\end{equation}
where $q=0.707$ and $p=0.3$ are obtained from a fit to numerical simulations.
 A similar model exists for ellipsoidal collapse \cite{Sheth:2001dp}.
\begin{eqnarray}\label{eq:Multibias1}
b_1   &=&1+\average{\frac{(q\nu-1)}{\delta_c}+\frac{2 p}{\delta_c\left(1+(q\nu)^p\right)}}\,,\\
b_2   &=&\frac{8}{21}\left(b_1   -1\right)+\average{\frac{4\left(p^2+ \nu p q\right)-(q\nu-1)\left(1+(q\nu)^p\right)-2p}
{\delta^2_c \left(1+(q\nu)^p\right)}
 +\frac{1}{\delta_c^2}\left((q\nu)^2 - 2q\nu -1\right)}\,
\label{eq:Multibias2}\\
b_{3} &=& -\frac{236}{189}\left( b_{1}-1\right) -\frac{13}{7} \left(b_{2}- \frac{8}{21}\left( b_{1}-1\right)\right)
+  \bigg<-\frac{\left(3 + 3 \nu q + 3 \nu^2 q^2 - \nu^3 q^3\right)}{\delta_c^3}
\\ \nonumber &&
+ \frac{\left(8 p^3 + 12 p^2 \left( 1 +\nu q\right) + p \left( 6 \nu^2 q^2 -2 \right) \right) }{\delta_c^3 \left( 1 + 1+ (\nu q)^p\right)} 
+6\frac{\left(1+ 2\nu q -\nu^2 q^2\right)}{\delta_c^3} - 24 \frac{\left( p^2 + \nu pq\right)}{\delta_c^3\left( 1+ (\nu q)^p\right)}
\\ \nonumber &&
 -4\frac{(1-\nu q)}{ \delta_c^3} + 8 \frac{p}{\delta_c^3\left( 1+ (q\nu)^p\right)}\bigg>_{\rm M}\,,
\label{eq:Multibias3}
\end{eqnarray}
 The HI bias parameters used in the paper were obtained by averaging over the halo bias parameters according to
\begin{equation}\label{eq:averagedef}
X  (z,{\x})=\average{X_h(z,{\x})}= \frac{\int_{M_{\rm{min}}}^{M_{\rm{max}}} \d M\left[ X_{h}(z,{\x},M)M_{\HI}  (M) n_h(z,{\x},M) \right] }{\int_{M_{\rm{min}}}^{M_{\rm{max}}} \d M\left[ M_{\HI}(M)
n_h(z,{\x},M)\right],
}
\end{equation}
where
$M_{\rm{min}}$ and $M_{\rm{max}}$ are the lower and upper limits of masses, which are related to the limits of circular velocity of galaxies that could house HI.  These are obtained from the circular velocity constraint
\begin{equation}\label{eq:velvsmass}
v_{\text{circ}} = 30 \sqrt{1+z} \left(\frac{M}{10^{10}M_{\odot}}\right)^{1/3} ~{\rm km\,s}^{-1}\,,
\end{equation}
We  assumed that only halos with circular velocities between $30 - 200\,$kms$^{-1}$ are able to host HI. This range of circular velocity is motivated by observation \cite{Bull:2014rha}. We adopt a simple polynomial fitting function for the HI mass function. This choice is based on the results from an  N-body simulation for the  HI mass function \cite{Santos:2015gra}
\begin{equation}
M_{\HI}(M) = C M^{0.6}\,,
\end{equation}
Here, the normalization factor is chosen to match the measurement of $\Omega_{\HI}$ at $z =0.8$ \cite{Switzer:2013ewa}.  We compute each of the bias parameters  in figure \ref{fig:biasparameters}.

The $\Omega_{\HI}$ at any redshift may also be obtained from the halo model 
\begin{eqnarray}
\Omega_{\HI}(z)\equiv \dfrac{1}{(1+z)^2} \frac{\rho_{\HI}(z)}{\rho_{c,0}}\,,
\end{eqnarray}
where $\rho_{c,0}$ is the homogeneous matter density today and $\rho_{\HI}$ is the density of the HI atoms
\begin{eqnarray}
\rho_{\HI}(z) = \int_{M_{\rm{min}}}^{M_{\rm{max}}} \d M \left[ M_{\HI}(M)
n_h(z,M)\right]\,.
\end{eqnarray}


\end{document}